%% file: main.tex
\documentclass{article}
\pdfoutput=1
\usepackage{graphicx}
\usepackage{booktabs}
\usepackage{authblk} 
\usepackage{amsmath}
\usepackage{amssymb}
\usepackage{xcolor}
\usepackage{multicol}
\usepackage{geometry}
\usepackage{float}
\usepackage{stmaryrd}
\usepackage{mathtools,empheq}
\usepackage{xr} 
\makeatletter
\newcommand*{\addFileDependency}[1]{
\typeout{(#1)}
\IfFileExists{#1}{}{\typeout{No file #1.}}
}\makeatother

\newcommand*{\myexternaldocument}[1]{%
\externaldocument{#1}%
\addFileDependency{#1.tex}%
\addFileDependency{#1.aux}%
}
\myexternaldocument{supplementary_material}
\usepackage[hidelinks]{hyperref}
\usepackage[capitalise]{cleveref}
\usepackage[ruled,vlined]{algorithm2e}

\usepackage[numbers,sort&compress]{natbib}
\usepackage[acronym]{glossaries}
\makeglossaries
\newacronym{CZM}{CZM}{cohesive zone model}
\newacronym{TSL}{TSL}{traction-separation law}
\newacronym{NSN}{NSN}{nonsmooth Newmark}
\newacronym{MLCP}{MLCP}{mixed linear complementarity problem}
\newacronym{LCP}{LCP}{linear complementarity problem}
\newacronym{QP}{QP}{quadratic programming}
\newacronym{SPSD}{SPSD}{symmetric positive semidefinite}
\newacronym{SPD}{SPD}{symmetric positive definite}
\newacronym{RVE}{RVE}{representative volume element}
\newacronym{CFL}{CFL}{Courant-Friedrichs-Lewy}
\newacronym{SDOF}{SDOF}{single degree-of-freedom}
\newacronym{MJ}{MJ}{Moreau-Jean}
\newacronym{CDL}{CDL}{CD-Lagrange}
\newacronym{LEO}{LEO}{Low Earth Orbit}
\newacronym{FEM}{FEM}{Finite Element Method}
\newacronym{SPH}{SPH}{Smoothed Particle Hydrodynamics}
\newacronym{XFEM}{XFEM}{eXtended Finite Element Method}
\newacronym{CD}{CD}{Central Difference}
\newacronym{NSCD}{NSCD}{Nonsmooth Contact Dynamics}
\newacronym{DOF}{DOF}{degree-of-freedom}
\newacronym{SBM}{SBM}{NASA Standard Breakup Model}
\newacronym{MEP}{MEP}{Maximum Entropy Principle}
\newacronym{HVI}{HVI}{hypervelocity impacts}
\newacronym{CCDF}{CCDF}{complementary cumulative distribution function}
\newacronym{DG}{DG}{discontinuous Galerkin}

\title{The semi-explicit nonsmooth Newmark time integrator for robust unilateral contact in dynamic fragmentation simulations}

\author[1]{Thibault Ghesquière-Diérickx}
\author[1]{Guillaume Anciaux\thanks{Corresponding author: \href{mailto:guillaume.anciaux@epfl.ch}{guillaume.anciaux@epfl.ch}}}
\author[2]{Vincent Acary}
\author[1]{Jean-François Molinari}
\affil[1]{Institute of Civil Engineering, Institute of Materials Science and Engineering, École Polytechnique Fédérale de Lausanne (EPFL), Lausanne, Switzerland}
\affil[2]{INRIA, Université Grenoble Alpes, Grenoble, France}

\date{}

\begin{document}

\maketitle

\begin{abstract}

    \noindent
    Numerical simulations of solids undergoing dynamic fragmentation, a problem characterized by dynamic fracture and dense contacts, require accurately capturing the transition from a solid continuum to a collection of interacting fragments. We use the finite-element method with the extrinsic cohesive zone model for fracture. For contact, conventional penalty-based methods often exhibit numerical instabilities in dynamic collision-rich settings. To address this, we adapt and validate a novel semi-explicit time-integration scheme: the Nonsmooth Newmark (NSN) method for unilateral contact. Based on the Nonsmooth Contact Dynamics (NSCD) method, this formulation strongly enforces contact constraints at the velocity level. Within this scheme, the bulk dynamics are non-impulsive and integrated explicitly with second-order accuracy, and the fracture model allows displacement discontinuities and integrates the cohesive softening explicitly. Contact, treated rigorously as nonsmooth, is integrated implicitly with first-order accuracy. Benchmark tests demonstrate that the NSN scheme achieves accuracy comparable to established nonsmooth methods, such as the semi-explicit CD-Lagrange and implicit Moreau-Jean schemes. Moreover, it outperforms penalty-based approaches by orders of magnitude. Although the NSN method incurs a higher per-step computational cost, its enhanced stability allows for significantly larger time steps. Consequently, for 1D benchmarks, overall computational efficiency is comparable to or better than that of purely explicit approaches. We applied this framework to 1D fragmentation under free and confined expansion. Results reveal that confinement shifts the fracture energy budget from local fragment kinetic energy to larger-scale global system kinetic energy. Additionally, we found, counterintuitively, that, compared to fully elastic contact, adding contact dissipation reduces fracture energy yet increases the final fragment count. It occurs because such dissipation reduces the vibration within damaged fragments, allowing cleaner stress-wave propagation and better damage localization, driving cracks to full separation rather than distributing damage. These results establish the NSN scheme as a robust tool for generating high-fidelity fragmentation statistics.
    
    \vspace{0.5cm}
    \noindent\textbf{Keywords:} cohesive zone model; semi-explicit time integration; nonsmooth contact dynamics; dynamic fragmentation; penalty methods; contact mechanics; finite element method; nonsmooth newmark

\end{abstract}



\input{01_introduction}
\input{02_theoretical_and_numerical_framework}
\input{03_numerical_validation}

\input{04_application_to_dynamic_fragmentation}
\input{05_conclusion}

\subsection*{Acknowledgements}
This research was funded by the Swiss National Science Foundation (SNSF) [Grant N° 212935].

\appendix
\onecolumn
\include{06_appendix}

\twocolumn
\bibliography{SWPaper}

\end{document}

%% file: 01_introduction.tex
\section{Introduction}

\medskip\noindent
Systems characterized by dense, simultaneous contacts and impacts are omnipresent in both nature and engineering, spanning applications from granular flows and geophysical hazards such as rock avalanches to hypervelocity impacts, fragmentation, and mechanical assemblies. Among these, dynamic fracture is especially demanding: the contact interfaces emerge dynamically as surfaces are created, when the solid continuum undergoes rapid topological evolution, shattering into numerous discrete fragments. This process couples stress‑wave–driven crack nucleation, growth, branching, and coalescence \citep{grady1985mechanisms, grady2007fragmentation, ramesh2015review, hild2015characteristic} with frequent, often simultaneous impacts between crack faces and fragments. This process of dynamic fragmentation occurs across vastly different scales---from the astronomical fragmentation of supernovae and planetary bodies \citep{abarzhi2019supernova}, to fragmentation of spacecraft in orbit \citep{liou2013debrisat,cowardin2023updates,murray2019analysis,olivieri2023analysis}, and molecular DNA fragmentation under ion irradiation \citep{gudowska2009distribution}. Such problems demand formulations capable of capturing both microsecond-scale fracture mechanics and long-term fragment dispersal. From a numerical perspective, accurately resolving this dual-timescale constraint requires millions of tiny integration steps, making numerical robustness a primary concern. Without a rigorous treatment of the nonsmooth mechanics associated with unilateral constraints and frictional laws that induce sharp discontinuities in velocity and forces, numerical artifacts can easily mask the underlying physics. The fidelity of the prediction of fragment kinematics depends, therefore, fundamentally on how the solid deformation, the fracture process, and contact interaction are discretized both in space and in time.

\medskip\noindent
Within the \acrfull{FEM} framework, fracture models are commonly grouped into diffuse and sharp‑interface approaches. Diffuse models such as phase‑field or lip-field \citep{miehe2010phase, francfort1998revisiting,chevaugeon2022lipschitz}, and related gradient‑damage formulations represent cracks by a continuous damage field and are very robust for complex crack paths, but do not provide a geometrically sharp interface for contact and fragment tracking. Sharp methods, including the \acrfull{XFEM} \citep{moes1999finite}, the \acrfull{CZM} \citep{dugdale1960yielding,barenblatt1962mathematical,xu1994numerical}, and \acrfull{DG} methods combined with cohesive interfaces \citep{radovitzky2011scalable,nguyen2014discontinuous}, introduce displacement discontinuities along well‑defined surfaces. We utilize the \acrshort{CZM} because this sharp-crack topology is critical for tracking fragment kinematics and resolving post-fracture collisions. This method enables the spontaneous initiation and propagation of multiple cracks within the solid by inserting zero-thickness interface elements, called cohesive elements, along the boundaries of continuum elements \citep{xu1994numerical}. We employ extrinsic \acrshort{CZM}s \citep{camacho1996computational}, characterized by the absence of an initial cohesive stiffness: elements are inserted dynamically only when a fracture criterion is met. This approach contrasts with intrinsic models, in which predefined cohesive zones require a high dummy stiffness ($k \to \infty$) to enforce continuity before failure, thereby introducing artificial compliance that alters wave speeds \citep{tijssens2000numerical}. By avoiding this initial stiffness entirely, the extrinsic approach maintains the exact elastic response of the bulk material until rupture. The robustness and scalability of explicit \acrshort{CZM} have been successfully demonstrated in large-scale 3D fragmentation problems \citep{camacho1996computational,ortiz1999finite,repetto2000finite,zhou2005cohesive,pandolfi1999finite}. In this work, we use the open-source and parallel \texttt{Akantu} FEM library \citep{richart2024akantu} that incorporates extrinsic cohesive elements.

\medskip\noindent
Accurately capturing contact and impact among crack faces and fragments is equally critical. Although contact mechanics are inherently nonsmooth, the contact constraints are typically regularized. Penalty-based methods approximate contact via artificial springs, preventing interpenetration between contacting bodies. These methods do not add additional \acrfull{DOF} and naturally fit within explicit time-integration schemes, e.g., the Newmark scheme ($\beta=0, \gamma=\frac12$), that are required to accurately capture the wave-driven fracture process in dynamic fragmentation. However, such contact methods have a fundamental limitation. While low contact stiffness values allow unphysical interpenetration, high contact stiffness values severely restrict the stable time-step size of the explicit scheme, thereby prohibiting the efficient simulation of large systems \citep{wriggers2006computational, laursen2003computational}. In addition, while the standard explicit Newmark scheme is symplectic and energy-conserving for smooth systems \citep{kane2000variational}, the interaction between fracture and regularized contact introduces severe nonsmoothness in the loading paths. It leads to a persistent numerical energy drift that can compromise the simulation's global stability. Our previous work \citep{ghesquiere2025stability} identified two potential mechanisms for this energy drift. As damage tends to zero, (i) the cohesive stiffness in tension diverges, and (ii) the sudden switch from tension (cohesion) to compression (contact) creates a discontinuous stiffness jump. The latter was identified as the main driver of global numerical instabilities. Previous efforts to mitigate these penalty-induced instabilities include bi-penalty formulations \citep{hetherington2012new, hetherington2013bipenalty, kolman2021bi}, mass redistribution schemes \citep{khenous2008mass}, and singular-mass adjustments \citep{renard2010singular}. These techniques remain partial fixes rather than fundamental resolutions of the underlying nonsmooth physics.

\medskip\noindent
To overcome the instabilities of penalty-based methods, we build upon the \acrfull{NSCD} framework \citep{moreau1988unilateral,jean1992unilaterality,jean1999non,moreau1999numerical}. This approach strongly enforces the Signorini gap constraints at the velocity level and allows instantaneous velocity jumps upon impact. Standard ``event-driven'' nonsmooth solvers---which precisely locate every collision time by adapting the time step---are unsuitable for dynamic fragmentation due to the high density of impacts. Instead, a robust ``time-stepping'' approach is required \citep{acary2008numerical}. Unlike regularized approaches that rely on single-valued force laws, the model employs set-valued force laws (solved as a \acrfull{MLCP} at each step) to determine contact impulses and post-impact velocities, thereby eliminating artificial compliance and associated energy drifts. Applying this logic to dynamic fragmentation requires a careful consideration of computational cost. While recent advancements have proposed a fully monolithic, purely implicit nonsmooth framework for \acrshort{CZM} \citep{collins2022formulation, collins2025formulation}, such an approach becomes computationally prohibitive for large-scale simulations. It is primarily due to the significant expansion of the system's dimensionality through the introduction of Lagrange multipliers, as well as the challenges in parallelizing the solvers required to resolve the globally coupled systems. 

\medskip\noindent
After identifying contact as the primary driver of instability in regularized approaches, we aim to apply the implicit, nonsmooth treatment exclusively to contact surfaces, while retaining explicit integration for the bulk material and the fracture process. To that end, we adopt the \acrfull{NSN} approach \citep{chen2013nonsmooth} and adapt it to such an explicit-implicit scheme. This targeted approach aims to contain computational cost: \acrshort{NSN} solves only contact constraints, with cost scaling with the active contact set rather than the total system size. Crucially, the resulting \acrfull{LCP} admits a convex \acrfull{QP} formulation, enabling efficient, parallelizable implementations of the contact solver. 

\medskip\noindent
Beyond validating the new scheme against existing nonsmooth and penalty-based approaches, we present a dynamic fragmentation case study that targets a particularly challenging regime of dynamic fragmentation: post-fracture collisions between fragments in a confined environment. The standard 1D benchmark of dynamic fragmentation---the free expansion of a bar or Mott expanding ring---recovers classical trends in fragment size statistics and energy dissipation, validating the fracture model and its coupling with the \acrshort{NSN} scheme. In contrast, enclosing fragments within rigid boundaries reveals how elastic fragment-wall and fragment-fragment collisions redistribute energy at the system scale, shifting the fracture energy budget from the local, fragment-scale kinetic energy \citep{grady1982local}, to the global one. We further test the model by adding contact dissipation, i.e., inelastic collisions, and investigate how much energy is dissipated through fracture or through contact. These confined 1D tests stress‑test the \acrshort{NSN} contact formulation with dense, simultaneous contact events. The observed stability and energy control, achieved without the prohibitive cost of global implicit solves, demonstrate that the method can preserve physical fidelity in dense-contact scenarios, a key requirement for extension to orbital‑debris–like scenarios, where frequent fragment–fragment interactions drive secondary fragmentation.

\medskip\noindent
The remainder of this paper is organized as follows. Section \ref{sec:theoretical_and_numerical_framework} details the theoretical formulation of the semi-explicit \acrshort{NSN} scheme and its algorithmic implementation as a \acrfull{QP} problem. It also introduces a modified cohesive \acrfull{TSL} that prevents prohibitively large cohesive stiffness, allowing for larger stable time steps. Section \ref{sec:numerical_validation} presents a numerical validation of the proposed framework in terms of stability and efficiency. We study three simple benchmarks: the bouncing ball problem, the impact of a bar on a rigid wall, and the impact of a damaged bar on a rigid wall. These scenarios compare the \acrshort{NSN} scheme with existing nonsmooth approaches (the implicit \acrfull{MJ} \citep{moreau1999numerical} and the semi-explicit \acrfull{CDL} \citep{fekak2017new}) and with the penalty-based approach. Finally, Section \ref{sec:application_to_dynamic_fragmentation} applies the framework to 1D dynamic fragmentation. By comparing free and confined expansion, we demonstrate the framework's readiness to handle and dense contact scenerios involving interacting cohesive cracks and complex multi-body interactions. This benchmark serves as a critical step toward higher dimensions. Section \ref{sec:conclusion} concludes the paper and outlines future research directions.

%% file: 02_theoretical_and_numerical_framework.tex
\section{Theoretical and numerical framework}
\label{sec:theoretical_and_numerical_framework}

The following section details the theoretical formulation of the semi-explicit \acrshort{NSN} scheme and its algorithmic implementation as a \acrshort{QP} problem. This framework is designed to handle elasticity, cohesive fracture, and unilateral contact. To ensure numerical stability and computational efficiency, we also introduce a modified cohesive \acrshort{TSL} that prevents prohibitively large cohesive stiffness.

\subsection{Variational formulation}
\label{sec:var_form}

\begin{figure}[h!]
    \centering
    \includegraphics[scale=1]{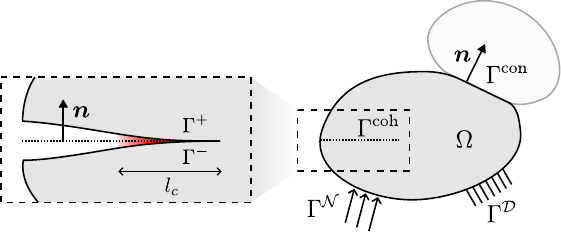}
    \caption{Geometry of the problem: Elastic body with Neumann boundary conditions on $\Gamma^\mathcal{N}$, Dirichlet boundary conditions on $\Gamma^\mathcal{D}$, cohesive interfaces on $\Gamma^\text{coh}$ and unilateral contact on $\Gamma^\text{con}\cup \Gamma^\text{coh}$, both split into $\Gamma^+$ and $\Gamma^-$}
    \label{fig:geometry}
\end{figure}

\paragraph{Geometry of the problem.} We consider an elastic body, Figure~\ref{fig:geometry}, occupying a reference configuration $\Omega \subset \mathbb{R}^d$ ($d=1,2,3$). The current position is $\boldsymbol x = \boldsymbol{X} + \boldsymbol u(\boldsymbol X,t)$, where $\boldsymbol X \in \Omega$ denotes the material point in the reference configuration and $\boldsymbol u(\boldsymbol X,t)$ the displacement field at time $t \in [0, T]$. The boundary $\partial \Omega$ is decomposed into two disjoint parts
\begin{equation}
    \partial \Omega = \Gamma^\mathcal{D} \cup \Gamma^\mathcal{N},
\end{equation}
where Dirichlet and Neumann boundary conditions are prescribed on $\Gamma^\mathcal{D}$ and $\Gamma^\mathcal{N}$, respectively. To account for cohesive fracture, we introduce a set of internal interfaces $\Gamma^\text{coh} \subset \overline{\Omega}$, with $\overline{\Omega} = \Omega \cup \partial\Omega$ the closure of the domain, along which the material may separate. Once an interface is created, two material points initially coinciding at $\boldsymbol X \in \Gamma^\text{coh}$ split into two distinct current positions $\boldsymbol x^\pm$ on the two faces $\Gamma^\pm$ of the interface. The displacement field may then be discontinuous, and we define the displacement jump as
\begin{equation}
  \boldsymbol{\delta} := \llbracket \boldsymbol{u} \rrbracket 
  = \boldsymbol{u}^+ - \boldsymbol{u}^-.
\end{equation}
Let $\boldsymbol n$ denote the unit normal to the interface, oriented from the ``$-$'' side to the ``$+$'' side. We decompose the displacement jump as
\begin{equation}
    \delta_n := \boldsymbol{\delta}\cdot\boldsymbol n,
    \qquad
    \boldsymbol{\delta}_t := ( \mathbf I - \boldsymbol n \otimes \boldsymbol n) \boldsymbol{\delta},
\end{equation}
so that $\delta_n$ is a scalar normal opening and $\boldsymbol{\delta}_t$ is the tangential (sliding) vector.

\medskip\noindent
In addition to cohesive fracture, the body may come into frictionless unilateral contact either (i) between the two faces of a cohesive interface $\Gamma^\text{coh}$ or (ii) between the external boundary $\Gamma^\text{con} \subset \partial \Omega$ and a rigid obstacle or another body. The normal gap function $g(\boldsymbol u)$ measures the signed distance between the current configuration of a point on the contact surface and the obstacle along the outward unit normal $\boldsymbol n$. On cohesive interfaces, we have $g(\boldsymbol u) = \delta_n$. The unilateral contact constraint requires
\begin{equation}
    g(\boldsymbol u) \ge 0 
    \qquad \text{on } \Gamma^\text{con} \cup \Gamma^\text{coh},
\end{equation}
which excludes penetration of the obstacle and of the cohesive faces.

\paragraph{Principle of virtual power with cohesion and contact.}

Let $\boldsymbol v$ be an admissible virtual velocity field and $\dot{\boldsymbol\varepsilon}(\boldsymbol v)$ its associated virtual strain rate. On cohesive interfaces, the virtual jump rate is
\begin{equation}
    \dot{\boldsymbol\delta}(\boldsymbol v) 
    := \llbracket \boldsymbol v \rrbracket
    = \boldsymbol v^+ - \boldsymbol v^- .
\end{equation}
The cohesive traction $\boldsymbol t^{\mathrm{coh}}$ is defined as the traction acting on the ``$+$'' face of the interface, exerted by the ``$-$'' face. The internal virtual power associated with bulk deformation and cohesive separation is
\begin{equation}
    \mathcal P^{\mathrm{int}}(\boldsymbol v)
    = - \int_\Omega \boldsymbol\sigma : \dot{\boldsymbol\varepsilon}(\boldsymbol v)\, dV
      + \int_{\Gamma^\text{coh}} 
            \boldsymbol t^{\mathrm{coh}} \cdot \dot{\boldsymbol\delta}(\boldsymbol v)\, dS,
\end{equation}
where $\boldsymbol\sigma$ is the Cauchy stress. Additional internal variables and their conjugate forces may be introduced. In the next section, we add damage through a thermodynamically consistent framework. The external and inertial virtual powers read
\begin{equation}
    \mathcal P^{\mathrm{ext}}(\boldsymbol v)
    = \int_\Omega \rho\,\boldsymbol b \cdot \boldsymbol v\, dV
      + \int_{\Gamma_N} \boldsymbol f^{\mathrm{ext}} \cdot \boldsymbol v\, dS,
\qquad
    \mathcal P^{\mathrm{in}}(\boldsymbol v)
    = \int_\Omega \rho\, \ddot{\boldsymbol u} \cdot \boldsymbol v\, dV.
\end{equation}

\medskip\noindent
The \emph{primal} statement of the principle of virtual power is:
\begin{equation}
    \mathcal P^{\mathrm{in}}(\boldsymbol v)
    = \mathcal P^{\mathrm{ext}}(\boldsymbol v) + \mathcal P^{\mathrm{int}}(\boldsymbol v)
    \qquad 
    \forall\, \boldsymbol v \in \mathcal V_0,
\end{equation}
subject to the unilateral constraint $g(\boldsymbol u)\ge 0$. The space of admissible virtual velocities is
\begin{equation}
    \mathcal V_0 
    = \bigl\{\, \boldsymbol v \in H^1(\Omega \setminus \Gamma^\text{coh})^d \;\big|\; 
        \boldsymbol v = \mathbf 0 \ \text{on } \Gamma^\mathcal{D} \,\bigr\}.
\end{equation}

\medskip\noindent
To avoid handling the geometric constraint $g(\boldsymbol u)\ge 0$ directly in the primal space, we introduce a nonnegative Lagrange multiplier $\lambda \ge 0$ on $\Gamma^\text{con} \cup \Gamma^\text{coh}$, representing the normal contact pressure. The virtual power associated with the contact then reads
\begin{equation}
    \mathcal P^{\mathrm{con}}(\boldsymbol v;\lambda)
    = - \int_{\Gamma^\text{con} \cup \Gamma^\text{coh}}
        \lambda\, \dot g(\boldsymbol v)\, dS,
\qquad
    \dot g(\boldsymbol v) = (\boldsymbol v^+ - \boldsymbol v^-)\cdot\boldsymbol n .
\end{equation}
The augmented virtual power balance is therefore
\begin{equation}
    \mathcal P^{\mathrm{in}}(\boldsymbol v)
    = \mathcal P^{\mathrm{ext}}(\boldsymbol v) + \mathcal P^{\mathrm{int}}(\boldsymbol v)
    + \mathcal P^{\mathrm{con}}(\boldsymbol v;\lambda)
    \qquad \forall\, \boldsymbol v \in \mathcal V_0.
\end{equation}
Together with the kinematic constraint and dual admissibility, the primal-dual problem takes the form of the classical Signorini-Fichera complementarity conditions:
\begin{equation}
    g(\boldsymbol u) \ge 0, 
    \qquad \lambda \ge 0,
    \qquad \lambda\,g(\boldsymbol u)=0
    \quad \text{on } \Gamma^\text{con} \cup \Gamma^\text{coh},
\end{equation}
compactly written as $0 \le g(\boldsymbol u) \perp \lambda \ge 0$ (where the symbol $\perp$ indicates orthogonality/complementarity, i.e., the product $g\lambda = 0$) on $\Gamma^\text{con} \cup \Gamma^\text{coh}$. Using integration by parts in the bulk and along cohesive and contact surfaces, and assuming sufficient smoothness of the displacement and stress fields (e.g., $\boldsymbol u\in C^2(\Omega\setminus\Gamma^\text{coh})$ and $\boldsymbol\sigma\in C^1(\Omega\setminus\Gamma^\text{coh})$), as well as Lipschitz continuity of the boundaries, the primal–dual virtual power statement yields the following strong form
\begin{equation}
    \begin{cases}
        \nabla \cdot \boldsymbol\sigma + \rho\,\boldsymbol b = \rho \ddot{\boldsymbol u} 
            & \text{in } \Omega \setminus \Gamma^\text{coh}, \\
        \boldsymbol u = \bar{\boldsymbol u} 
            & \text{on } \Gamma^\mathcal{D}, \\
        \boldsymbol\sigma \boldsymbol n 
            = \boldsymbol f^{\mathrm{ext}} 
            & \text{on } \Gamma^\mathcal{N}, \\
        \boldsymbol\sigma \boldsymbol n 
            = -\lambda\,\boldsymbol n 
            & \text{on } \Gamma^\text{con}, \\
        \boldsymbol\sigma^\pm \boldsymbol n 
            = \boldsymbol t^{\mathrm{coh}} - \lambda \boldsymbol n, 
            & \text{on } \Gamma^\text{coh}, \\
        0 \le g(\boldsymbol u) \ \perp\ \lambda \ge 0 
            & \text{on } \Gamma^\text{con} \cup \Gamma^\text{coh}.
    \end{cases}
\end{equation}

\medskip\noindent
The constitutive structure of $(\boldsymbol\sigma,\boldsymbol t^{\mathrm{coh}})$ will be specified in the next section through free energies, reversible state laws, and irreversible evolution equations, including a Camacho--Ortiz-type cohesive model with damage.

\paragraph{Free energy and reversible state laws.} We consider the small-strain tensor $\boldsymbol\varepsilon$ in the bulk and the displacement jump $\boldsymbol{\delta}$ across cohesive interfaces as reversible state variables. To account for irreversible separation, a scalar internal damage variable $d \in (0,1]$ is introduced on $\Gamma^\text{coh}$. The total free energy of the system reads
\begin{equation}
    \Psi(\boldsymbol\varepsilon, \delta, d) = \int_\Omega \Psi^e(\boldsymbol\varepsilon)\,dV + \int_{\Gamma^\text{coh}} \Psi^\text{coh}(\delta, d)\,dS,
\end{equation}
with $\Psi^e$ the bulk elastic energy density and $\Psi^\text{coh}$ the cohesive free energy density. Assuming linear elasticity of the bulk, the elastic energy density is given by
\begin{equation}
    \Psi^e(\boldsymbol\varepsilon) = \tfrac12\, \boldsymbol\varepsilon : \mathbf C : \boldsymbol\varepsilon,
\end{equation}
with $\mathbf C$ the fourth-order elasticity tensor. The reversible stress follows from the energy derivative:
\begin{equation}
    \boldsymbol\sigma 
    = \partial_{\boldsymbol\varepsilon} \Psi^e(\boldsymbol\varepsilon) = \mathbf C : \boldsymbol\varepsilon.
\end{equation}
For the cohesive interfaces, we adopt a potential-based \acrfull{CZM}. The mixed-mode effective measure characterizes the opening
\begin{equation}
    \delta = \sqrt{\delta_n^2 + \beta^2\|\boldsymbol{\delta}_t\|^2},
\end{equation}
where the dimensionless parameter $\beta$ weighs the normal and tangential contributions. To reproduce the linear irreversible \acrfull{TSL} of Camacho and Ortiz~\citep{camacho1996computational} within a thermodynamically consistent framework, we use the damage variable $d$ and define the cohesive free energy density as
\begin{equation}
    \Psi^\text{coh}(\delta, d) 
    = \frac12 \frac{1-d}{d}\frac{\sigma_c}{\delta_c} \delta^2,
\end{equation}
where $\sigma_c$ is the local cohesive strength and $\delta_c$ is the critical opening beyond which complete decohesion occurs. For a fixed $d$, this free energy generates a linear traction-separation response with secant stiffness
\begin{equation}
     k(d) = \frac{1-d}{d} \frac{\sigma_c}{\delta_c},
     \label{eq:secant_stiffness}
\end{equation}
which coincides with the Camacho–Ortiz unloading/reloading stiffness. The associated cohesive traction follows from the chain rule:
\begin{equation}
    \boldsymbol t^{\mathrm{coh}} 
    = \partial_{\boldsymbol{\delta}} \Psi^\text{coh}(\delta, d) 
    = \frac{1-d}{d} \frac{\sigma_c}{\delta_c} \frac{1}{\delta}
      \left(\delta_n \boldsymbol n + \beta^2 \boldsymbol{\delta}_t \right). 
\end{equation}

\paragraph{Dissipation and irreversible processes.}

We consider an isothermal setting in which the material response is decomposed into reversible and irreversible processes. The reversible response is fully characterized by the free energy $\Psi$ introduced above. The bulk is assumed to be purely elastic, such that the mechanical dissipation vanishes in $\Omega$:
\begin{equation}
    \mathcal D^{\mathrm{bulk}} 
    = \boldsymbol{\sigma} : \dot{\boldsymbol\varepsilon} - \dot{\Psi}^e 
    = 0.
\end{equation}
Consequently, all dissipation is localized on the cohesive interfaces. On $\Gamma^\text{coh}$, the irreversible process is governed by the scalar internal variable $d$, which quantifies the loss of cohesive stiffness (damage). Although the physical fracture process involves the creation of new surfaces---a state of stored potential energy---the mechanical mechanism of decohesion is fundamentally irreversible ($\dot d \ge 0$). Because the energy required to break atomic bonds cannot be recovered as elastic potential, this structural change is treated as mechanical dissipation within the continuum framework, governed by the local Clausius-Duhem inequality:
\begin{equation}
    \mathcal D^{\mathrm{coh}}
    = \boldsymbol t^{\mathrm{coh}}\cdot\dot{\boldsymbol\delta}
      - \dot\Psi^\text{coh}(\delta,d)
    \;\ge\; 0.
\end{equation}
Using the state law $\boldsymbol t^{\mathrm{coh}}=\partial_{\boldsymbol\delta}\Psi^\text{coh}$ and applying the chain rule to the free energy $\Psi^\text{coh}$, the dissipation reduces to a concise form:
\begin{equation}
    \mathcal D^{\mathrm{coh}}
    = -\,\partial_d\Psi^\text{coh}(\delta,d)\,\dot d = G\,\dot d \;\ge\; 0.
\end{equation}
The irreversible damage process is therefore driven by the thermodynamic force $G$ conjugate to damage $d$, which is identified as the cohesive energy release rate:
\begin{equation}
    G = -\,\partial_d\Psi^\text{coh}(\delta,d) = \frac12 \frac{\sigma_c}{\delta_c} \frac{\delta^2}{d^2} \geq 0.
\end{equation}
At complete decohesion ($d=1, \delta=\delta_c$), this energy release rate reaches its critical fracture toughness value $G_c = \sigma_c \delta_c / 2$. To ensure $\mathcal{D}^\text{coh} \ge 0$, the damage evolution must satisfy $\dot d \ge 0$. This irreversibility condition is typically enforced through a rate-independent evolution law based on a loading function 
\begin{equation}
    f(\delta, d) := \frac{\delta}{\delta_c} - d \le 0 \quad \Leftrightarrow \quad \frac{\delta}{\delta_c} \le d.
\end{equation}
From the Kuhn–Tucker conditions associated with this constraint, we have
\begin{equation}
    \dot d \ge 0,
    \qquad
    f(\delta, d) \le 0,
    \qquad
    \dot d\,f(\delta, d) = 0,
    \label{eq:decohesion_complementarity}
\end{equation}
which ensures that damage grows only when the loading function reaches zero, i.e., when $\delta = d\,\delta_c$. These conditions are equivalent to the classical Camacho–Ortiz decohesion criterion that introduces the history-dependent evolution law
\begin{equation}
    \delta_{\max}(t) := \max_{0 \le \tau \le t} \delta(\tau),
    \qquad
    d(t) := \frac{\delta_{\max}(t)}{\delta_c},
    \label{eq:camacho_ortiz_irreversibility}
\end{equation}
ensuring that damage increases only when the current opening $\delta$ attains a new maximum.

\subsection{Space and time discretization}
\label{ss:space_and_time_discretization}

\paragraph{Space discretization.} The spatial discretization is carried out using standard finite elements. The displacement field is approximated as $\boldsymbol{u}(\boldsymbol{x},t) \approx \sum_{i=1}^n \mathbf{N}_i(\boldsymbol{x}) \mathbf{u}_i(t)$, where $\mathbf{N}_i(\boldsymbol{x})$ are the shape functions and $\mathbf{u}_i(t)$ are the nodal displacements, with duplicate nodes along $\Gamma^\text{coh}$ to capture displacement jumps. The contact pressure is discretized by piecewise constant Lagrange multipliers defined at the contact nodes on $\Gamma^\text{con} \cup \Gamma^\text{coh}$. Inserting these approximations into the virtual power principle, choosing nodal basis functions as test functions, and assembling over all elements and interface elements, we obtain the semi-discrete equation of motion
\begin{equation}
    \mathbf M \dot{\mathbf v}(t) + \mathbf K(\mathbf{d}) \mathbf u(t) = \mathbf f^{\mathrm{ext}}(t) + \mathbf H^\top \boldsymbol\lambda(t), \quad \mathbf{v}(t) = \dot{\mathbf u}(t).
    \label{eq:semi-discrete-motion}
\end{equation}
In this system, $\mathbf M \in \mathbb{R}^{n \times n}$ is the mass matrix. The matrix $\mathbf K \in \mathbb{R}^{n \times n}$ is the stiffness matrix that collects the bulk and cohesive contributions. It is explicitly state-dependent, as indicated by $\mathbf d$, which represents the history of cohesive damage variables. These variables are updated locally at the interface integration points to satisfy the irreversibility condition ($\dot d \ge 0$) of the damage process before the global matrix is assembled. The vector $\mathbf f^{\mathrm{ext}}(t) \in \mathbb{R}^{n}$ collects the external forces. The terms related to contact involve the matrix $\mathbf H \in \mathbb{R}^{m \times n}$ and the vector of nonnegative contact pressures $\boldsymbol\lambda(t) \in \mathbb{R}^{m}$, where $m$ is the number of contact nodes. The matrix $\mathbf H$ maps nodal displacements to normal gaps $\mathbf{g}(t) \in \mathbb{R}^m$ via
\begin{equation}
    \mathbf{g}(t) = \mathbf{H} \mathbf{u}(t).
    \label{eq:gap}
\end{equation}
The physical requirement of non-penetration is then enforced through the discrete complementarity conditions:
\begin{equation}
    0 \le \mathbf g(t) \ \perp\ \boldsymbol\lambda(t) \ge 0.
\end{equation}

\paragraph{Time discretization} 
Unilateral contact can induce discontinuities in the velocities of finite-dimensional dynamical systems. At such impact events, a contact force $\boldsymbol\lambda(t) \in L^1$ cannot produce an instantaneous change in velocity. To account for this impulsive behavior, the time discretization must be carefully designed \citep{moreau1999numerical}, and the reaction forces must be treated as vector-valued measures. The equations of motion~\eqref{eq:semi-discrete-motion} are rewritten in distributional form, an equality of measures, as
\begin{equation}
\mathbf M \, d\mathbf v(t) + \mathbf K \mathbf u(t)\,dt = \mathbf f^{\mathrm{ext}}(t)\,dt + \mathbf H^\top \, d\mathbf{i}(t).
\label{eq:distributional_motion}
\end{equation}
Because impacts cause velocity jumps, the velocity $\mathbf{v}$ is a function of bounded variation; its distributional derivative $d\mathbf{v}$ therefore comprises an absolutely continuous term and an atomic term supported on the countable set of collision times $\{t_c\}$. The contact impulse measure $d\mathbf{i}$ has the same structure.

\medskip\noindent
Equating the terms that are absolutely continuous with respect to the Lebesgue measure $dt$ yields the standard smooth equations of motion. The corresponding densities are the acceleration $\mathbf{\dot{v}}(t)$ for $d\mathbf{v}$ and the contact force $\boldsymbol\lambda(t)$ for $d\mathbf{i}$, giving
\begin{equation}
    \mathbf M \mathbf{\dot{v}}(t) + \mathbf K \mathbf u(t) = \mathbf f^{\mathrm{ext}}(t) + \mathbf H^\top \boldsymbol\lambda(t),
    \label{eq:smooth_motion}
\end{equation}
valid almost everywhere. Equivalently, equating the atomic terms concentrated on $\{t_c\}$ gives the momentum balance at impact. Their densities are defined with respect to the sum of Dirac measures $\sum_{c} \delta_{t_c}$, where $\delta_{t_c}$ is the Dirac measure centered at $t_c$. For $d\mathbf{v}$, this density is the velocity jump $\llbracket \mathbf{v} \rrbracket(t_c) = \mathbf{v}(t_c^+) - \mathbf{v}(t_c^-)$, and for $d\mathbf{i}$, it is the contact impulse $\mathbf{j}(t_c)\ge 0$. This yields the momentum balance at impact:
\begin{equation}
    \mathbf M \llbracket \mathbf{v} \rrbracket(t_c) = \mathbf H^\top \mathbf j(t_c).
    \label{eq:impact_balance}
\end{equation}
In this finite-dimensional setting, this impact equation is underdetermined ($n$ equations, $n+m$ unknowns, i.e., the post-impact velocities and the contact impulses). To close the system, we use Newton's impact law to relate the pre- and post-impact velocities. At $t_c$, the gap closes $\mathbf H\mathbf{u}(t_c)=0$ with negative relative velocity $\mathbf H\mathbf{v}(t_c^-)<0$, the impact law reads
\begin{equation}
    0 \le \mathbf j(t_c) \ \perp\ \mathbf H\mathbf v(t^+) + e\,\mathbf H\mathbf v(t^-) \ge 0,
\end{equation}
with $e \in [0,1]$ the coefficient of restitution. It phenomenologically captures the energy dissipation inherent in the collision, aggregating effects such as material plasticity, microcracking, acoustic emission, and heat generation into a constitutive relation.

\medskip\noindent
We can employ several time-stepping strategies to integrate the nonsmooth equations of motion. A widely used approach is the \emph{Moreau-Jean scheme} \citep{moreau1988unilateral,jean1992unilaterality,jean1999non,moreau1999numerical}, which consists of integrating the distributional form of the momentum balance over each time interval $[t_n,t_{n+1}]$. This integration yields a first-order-accurate method for the velocities, while the displacements are updated using the $\theta$-method. In contrast, the technique adopted here follows the \emph{nonsmooth Newmark scheme} introduced by Chen et al.~\citep{chen2013nonsmooth}. The key idea is an algorithmic split of the dynamics into two parts: a contribution $\boldsymbol{\widetilde\boxdot}$ associated with bulk deformation, external loading, and cohesive fracture, and a contribution $\boldsymbol{\widehat{\boxdot}}$ that gathers the entire contact reaction. By an abuse of terminology, we shall refer to the former as the ``smooth'' part and to the latter as the ``nonsmooth'' part.

\medskip\noindent
Following Chen et al. \citep{chen2013nonsmooth}, the velocity measure is split so that the smooth part $\dot{\widetilde{\mathbf{v}}}\,dt$ contains only the free-flight inertial and internal terms, while the nonsmooth remainder $d\widehat{\mathbf{v}}$ carries all contact-induced velocity changes. Thus the full contact measure $d\mathbf{i}$---comprising both the absolutely continuous contact forces (density $\boldsymbol{\lambda}(t)$) and the atomic impulses at impact times---is gathered in the nonsmooth side. This assignment differs from the Lebesgue decomposition introduced earlier, and yields the split dynamics
\begin{equation}
    \begin{cases}
        \mathbf{M} \dot{\widetilde{\mathbf{v}}}\,dt + \mathbf{K} \mathbf{u}(t)\,dt = \mathbf{f}^{\mathrm{ext}}(t)\,dt & \text{(Smooth dynamics)} \\
        \mathbf{M}\, d\widehat{\mathbf{v}} = \mathbf{H}^\top d\mathbf{i}(t) & \text{(Nonsmooth dynamics)}
    \end{cases}
\end{equation}
The smooth part is discretized using a second-order explicit Newmark-$\beta$ scheme (with $\beta=0$ and $\gamma=1/2$), while the nonsmooth part employs a first-order implicit Euler scheme. The discrete smooth acceleration $\dot{\widetilde{\mathbf{v}}}_{n+1}$ satisfies
\begin{equation}
    \mathbf{M} \dot{\widetilde{\mathbf{v}}}_{n+1} + \mathbf{K} \mathbf{u}_{n+1} = \mathbf{f}^{\mathrm{ext}}_{n+1}.
\end{equation}
The impulsive velocity increment $\widehat{\mathbf{v}}_{n+1}$ is obtained by approximating the integral of the nonsmooth measure over the time step $[t_n, t_{n+1}]$:
\begin{equation}
    \mathbf{M} \widehat{\mathbf{v}}_{n+1} \approx \mathbf{H}^\top \mathbf{p}_{n+1},
\end{equation}
where $\mathbf{p}_{n+1}$ is the total contact impulse accumulated during the time step, comprising both the integral of the contact forces and the sum of discrete impact impulses. To close the system, Newton's impact law is enforced over the active constraint set $\mathcal{A} = \{ i \in \{1, \ldots, m\}: g_i \le 0 \}$, where $\mathbf{g}$ is a gap prediction vector that identifies potential contact points. This prediction depends on the specific time-integration scheme and can be based on either the gap at the previous time step, $t_n$, or a predicted gap at $t_{n+1}$. We define $\mathbf{H}^\mathcal{A} \in \mathbb{R}^{|\mathcal{A}| \times n}$ as the submatrix of $\mathbf{H}$ restricted to the rows corresponding to the indices in $\mathcal{A}$. Similarly, $\mathbf{p}^{\mathcal{A}}_{n+1} \in \mathbb{R}^{|\mathcal{A}|}$ is the reduced impulse vector for these active constraints. The discretized impact law is then expressed as:
\begin{equation}
    \mathbf{0} \le \mathbf{p}^{\mathcal{A}}_{n+1} \ \perp\ \mathbf{H}^\mathcal{A} \mathbf{v}_{n+1} + e(\mathbf{H}^\mathcal{A} \mathbf{v}_n) \ge \mathbf{0}.
\end{equation}
By restricting the complementarity condition to $\mathcal{A}$, we ensure that impulses $p_{n+1,i}$ are identically zero for all inactive constraints $i \notin \mathcal{A}$. Following this discretization, the smooth displacements and velocities are updated using the acceleration variable $\mathbf{a} = \dot{\widetilde{\mathbf{v}}}$:
\begin{equation}
    \begin{cases}
        \widetilde{\mathbf{u}}_{n+1} = \mathbf{u}_n + \Delta t\,\mathbf{v}_n + \tfrac{\Delta t^2}{2}\,\mathbf{a}_n, \\
        \widetilde{\mathbf{v}}_{n+1} = \mathbf{v}_n + \tfrac{\Delta t}{2}(\mathbf{a}_n + \mathbf{a}_{n+1}).
    \end{cases}
\end{equation}
This smooth state serves as the ``free'' configuration to predict the active contact set $\mathcal{A}$. Specifically, we define the predicted gap as $\mathbf{g} = \mathbf{H} \widetilde{\mathbf{u}}_{n+1}$, and identify all indices $i$ such that $g_i \le 0$. To account for impulsive terms within $[t_n, t_{n+1}]$, the nonsmooth correction $\widehat{\mathbf{v}}_{n+1}$ is added to the smooth predictions:
\begin{equation}
    \begin{cases}
        \mathbf{u}_{n+1} = \widetilde{\mathbf{u}}_{n+1} + \tfrac{\Delta t}{2} \widehat{\mathbf{v}}_{n+1}, \\
        \mathbf{v}_{n+1} = \widetilde{\mathbf{v}}_{n+1} + \widehat{\mathbf{v}}_{n+1}.
    \end{cases}
\end{equation}
The resulting semi-explicit \acrfull{NSN} time integration scheme for unilateral contact with impacts is summarized below.

\begin{subequations}
    \begin{empheq}[left=\empheqlbrace]{align}
        &\mathbf{M} \mathbf{a}_{n+1} + \mathbf{K} \mathbf{u}_{n+1} = \mathbf{f}^{\mathrm{ext}}_{n+1} \label{eq:nsn_smooth_equ} \\
        &\mathbf{M} \mathbf{\widehat v}_{n+1} = \left(\mathbf{H}^\mathcal{A}\right)^\top \mathbf{p}^{\mathcal{A}}_{n+1} \label{eq:nsn_nsmooth_equ}\\
        &\mathbf{0} \le \mathbf{p}^{\mathcal{A}}_{n+1} \perp \mathbf{H}^\mathcal{A} \mathbf{v}_{n+1} + e \mathbf{H}^\mathcal{A}\mathbf{v}_{n} \ge \mathbf{0}, \quad \text{with } \mathcal{A} = \{ i \mid g_i(\widetilde{\mathbf{u}}_{n+1}) \le 0 \} \label{eq:nsn_contact}\\
        &\widetilde{\mathbf{u}}_{n+1} = \mathbf{u}_{n} + \Delta t \mathbf{v}_{n} + \tfrac{\Delta t^{2}}{2} \mathbf{a}_{n} \label{eq:nsn_smooth_displ}\\
        &\widetilde{\mathbf{v}}_{n+1} = \mathbf{v}_{n} + \tfrac{\Delta t}{2}(\mathbf{a}_{n} + \mathbf{a}_{n+1}) \label{eq:nsn_smooth_vel}\\
        &\mathbf{u}_{n+1} = \mathbf{\widetilde u}_{n+1} + \tfrac{\Delta t}{2} \mathbf{\widehat v}_{n+1} \label{eq:nsn_displ}\\
        &\mathbf{v}_{n+1} = \mathbf{\widetilde v}_{n+1} + \mathbf{\widehat v}_{n+1} \label{eq:nsn_vel}
    \end{empheq}
\end{subequations}

\subsection{Formulation as a constrained optimization problem}
Solving the system~\eqref{eq:nsn_smooth_equ}--\eqref{eq:nsn_vel} requires handling the coupling between the smooth and nonsmooth dynamics. To this end, we reformulate the problem as a \acrfull{LCP} restricted to the active set $\mathcal{A}$. With unknowns $\mathbf{a}_{n+1}$ and $\mathbf{p}^{\mathcal{A}}_{n+1}$, we rewrite the system to feed it to a dedicated solver.

\medskip\noindent
From the nonsmooth relation~\eqref{eq:nsn_nsmooth_equ}, we express the contact-induced velocity jump as
\begin{equation}
    \widehat{\mathbf v}_{n+1} = \mathbf M^{-1}\left(\mathbf H^{\mathcal{A}}\right)^{\!\top} \mathbf p^{\mathcal{A}}_{n+1}.
\end{equation}
Substituting this expression into the kinematic updates~\eqref{eq:nsn_displ}--\eqref{eq:nsn_vel} yields
\begin{subequations}
    \label{eq:kin_update_contact}
    \begin{align}
    \mathbf u_{n+1} &= \widetilde{\mathbf u}_{n+1} + \frac{\Delta t}{2}\,\mathbf M^{-1}\left(\mathbf H^{\mathcal{A}}\right)^{\!\top} \mathbf p^{\mathcal{A}}_{n+1}, \label{eq:kin_update_u} \\
    \mathbf v_{n+1} &= \mathbf v_n + \frac{\Delta t}{2}\bigl(\mathbf a_n + \mathbf a_{n+1}\bigr) + \mathbf M^{-1}\left(\mathbf H^{\mathcal{A}}\right)^{\!\top} \mathbf p^{\mathcal{A}}_{n+1}. \label{eq:kin_update_v}
    \end{align}
\end{subequations}
Inserting~\eqref{eq:kin_update_u}--\eqref{eq:kin_update_v} into the equilibrium~\eqref{eq:nsn_smooth_equ} and the impact law~\eqref{eq:nsn_contact} results in a mixed linear complementarity problem in the unknown pair $(\mathbf a_{n+1},\mathbf p^{\mathcal{A}}_{n+1})$:
\begin{subequations}
    \begin{align}
    &\mathbf M \mathbf a_{n+1} + \frac{\Delta t}{2}\,\mathbf K\mathbf M^{-1}\left(\mathbf H^{\mathcal{A}}\right)^{\!\top}\! \mathbf p^{\mathcal{A}}_{n+1} + \mathbf K\widetilde{\mathbf u}_{n+1} - \mathbf f_{n+1}^{\mathrm{ext}} = \mathbf 0, \label{eq:lcp_equilibrium} \\
    &\mathbf 0 \le \frac{\Delta t}{2}\mathbf H^{\mathcal{A}} \mathbf a_{n+1} + \mathbf H^{\mathcal{A}}\mathbf M^{-1}\left(\mathbf H^{\mathcal{A}}\right)^{\!\top}\! \mathbf p^{\mathcal{A}}_{n+1} + \mathbf H^{\mathcal{A}}\!\left(\frac{\Delta t}{2}\,\mathbf a_n + (1+e)\,\mathbf v_n\right) \perp \mathbf p^{\mathcal{A}}_{n+1} \ge \mathbf 0. \label{eq:lcp_complementarity}
    \end{align}
\end{subequations}
Eliminating the acceleration $\mathbf a_{n+1}$ from the equilibrium equation~\eqref{eq:lcp_equilibrium} and substituting the result into the complementarity condition~\eqref{eq:lcp_complementarity} yields a standard \acrfull{LCP} in the contact impulse alone:
\begin{equation}
    \mathbf 0 \le \mathbf W'\mathbf p^{\mathcal{A}}_{n+1} + \mathbf b \perp \mathbf p^{\mathcal{A}}_{n+1} \ge \mathbf 0,
    \label{eq:lcp}
\end{equation}
with the modified Delassus operator and the right-hand side given directly by
\begin{align}
    \mathbf W' &:= \mathbf H^{\mathcal{A}}\mathbf M^{-1}\!\left[\mathbf I - \frac{\Delta t^{2}}{4}\,\mathbf K\mathbf M^{-1}\right]\!\left(\mathbf H^{\mathcal{A}}\right)^{\!\top}, \\
    \mathbf b &:=\mathbf H^{\mathcal{A}}\left[(1+e)\,\mathbf v_n + \frac{\Delta t}{2}\,\mathbf a_n-\frac{\Delta t}{2}\,\mathbf M^{-1}\bigl(\mathbf K\widetilde{\mathbf u}_{n+1} - \mathbf f_{n+1}^{\mathrm{ext}}\bigr)\right]
    \label{eq:qp_operators}
\end{align}
Defining the contact-free velocity predictor
\begin{align}
    \mathbf v_{n+1}^{\mathrm{free}} := \mathbf v_n + \frac{\Delta t}{2}\,\mathbf a_n + \frac{\Delta t}{2}\,\mathbf M^{-1}\!\bigl(\mathbf f_{n+1}^{\mathrm{ext}} - \mathbf K\widetilde{\mathbf u}_{n+1}\bigr),
\end{align}
the vector $\mathbf b$ can be written compactly as $\mathbf b = \mathbf H^{\mathcal A}(\mathbf v_{n+1}^{\mathrm{free}} + e\mathbf v_n)$. Whenever $\mathbf W'$ is \acrfull{SPSD}, this \acrshort{LCP} is equivalent to the convex \acrfull{QP} over the active set:
\begin{equation}
    \min_{\mathbf p^{\mathcal{A}}_{n+1} \ge \mathbf 0} \frac12\left(\mathbf p^{\mathcal{A}}_{n+1}\right)^\top \mathbf W' \mathbf p^{\mathcal{A}}_{n+1} + \left(\mathbf p^{\mathcal{A}}_{n+1}\right)^\top \mathbf b.
    \label{eq:qp}
\end{equation}
As shown in Appendix~\ref{appendix:spd}, a sufficient condition for $\mathbf W'$ to be \acrshort{SPSD} is that the time step satisfies
\begin{equation}
    \Delta t \;\le\; \frac{2}{\omega_{\max}},
    \label{eq:stable_timestep}
\end{equation}
where $\omega_{\max}$ denotes the largest natural frequency of the system that the contact subspace excites. The standard central-difference stability bound can be used as a safe choice. The predictor–corrector semi-explicit \acrshort{NSN} scheme is summarized in Algorithm~\ref{alg:nsn_qp}.

\begin{algorithm}[h!]
    \DontPrintSemicolon
    \KwIn{$\mathbf{u}_0,\mathbf{v}_0,\mathbf{a}_0,\Delta t,e,t_{\mathrm{end}},
        \mathbf{M},\mathbf{K},\mathbf{H},
        \{\mathbf f^{\mathrm{ext}}_{n}\}_{n\ge 0}$}
    \KwOut{Trajectory $\{\mathbf{u}_n,\mathbf{v}_n,\mathbf{a}_n\}$}
    \caption{Predictor–corrector explicit \acrshort{NSN} with QP contact solve}
    \label{alg:nsn_qp}

    $t\gets 0$, $n\gets 0$\;

    \While{$t<t_{\mathrm{end}}$}{
        $t\gets t+\Delta t$, \quad $n\gets n+1$\;
        $\widetilde{\mathbf{u}}_{n+1} \gets \mathbf{u}_{n} + \Delta t\,\mathbf{v}_{n} + \tfrac{\Delta t^{2}}{2}\,\mathbf{a}_{n}$ \tcp*{\eqref{eq:nsn_smooth_displ}}
        Update $\mathbf{d}$ from $\widetilde{\mathbf{u}}_{n+1}$ and assemble $\mathbf K(\mathbf{d})$\;
        Gap prediction $\mathbf g_{n+1}\gets \mathbf{H}\widetilde{\mathbf{u}}_{n+1}$ \tcp*{\eqref{eq:gap}}
        Active set $\mathcal{A} = \{ i \mid g_i(\widetilde{\mathbf{u}}_{n+1})\le 0 \}$\;

        \If{$\mathcal A=\emptyset$}{
            No contact case: smooth update with $\mathbf{\widehat v}_{n+1}=\mathbf{0}$\;
            $\mathbf u_{n+1}\gets \widetilde{\mathbf u}_{n+1}$\tcp*{\eqref{eq:nsn_displ}} 
            $\mathbf a_{n+1}\gets \mathbf M^{-1}\!\big(\mathbf f^{\mathrm{ext}}_{n+1} -\mathbf K\,\mathbf u_{n+1}\big)$ \tcp*{\eqref{eq:nsn_smooth_equ}}
            $\mathbf v_{n+1}\gets
            \mathbf v_n+\tfrac{\Delta t}{2}\big(\mathbf a_n+\mathbf a_{n+1}\big)$ \tcp*{\eqref{eq:nsn_smooth_vel}}
        }
        \Else{
            Active contact case: solve QP for contact impulses\\
            Restrict $\mathbf H$ to the active set $\mathbf{H}^\mathcal{A}$\\


            Assemble operators $\mathbf W'$ and $\mathbf b$ \tcp*{\eqref{eq:qp_operators}}
            $\mathbf W' \gets \mathbf H^{\mathcal{A}}\mathbf M^{-1}\!\left[\mathbf I - \frac{\Delta t^{2}}{4}\,\mathbf K\mathbf M^{-1}\right]\!\left(\mathbf H^{\mathcal{A}}\right)^{\!\top}$ \;
            $\mathbf b \gets \mathbf H^{\mathcal{A}}\left[(1+e)\,\mathbf v_n + \frac{\Delta t}{2}\,\mathbf a_n-\frac{\Delta t}{2}\,\mathbf M^{-1}\bigl(\mathbf K\widetilde{\mathbf u}_{n+1} - \mathbf f_{n+1}^{\mathrm{ext}}\bigr)\right]$ \;

            Solve the QP:  $\mathbf p^\mathcal{A}_{n+1}\gets \arg\min_{\mathbf p\ge 0}\, \tfrac12\,\mathbf p^\top \mathbf W'\mathbf p +\mathbf p^\top \mathbf b$ \tcp*{\eqref{eq:qp}}

            Velocity correction: $\mathbf{\widehat{v}}_{n+1} \gets \mathbf{M}^{-1} \left(\mathbf{H}^\mathcal{A}\right)^\top \mathbf{p}^{\mathcal{A}}_{n+1}$  \tcp*{\eqref{eq:nsn_nsmooth_equ}}

            Nonsmooth update\;
            $\mathbf{u}_{n+1} \gets \mathbf{\widetilde{u}}_{n+1} + \frac{\Delta t}{2} \mathbf{\widehat{v}}_{n+1}$ \tcp*{\eqref{eq:nsn_displ}}
            $\mathbf{v}_{n+1} \gets \mathbf{\widetilde{v}}_{n+1} + \mathbf{\widehat{v}}_{n+1}$ \tcp*{\eqref{eq:nsn_vel}}
            $\mathbf{a}_{n+1} \gets \mathbf{M}^{-1} \left( \mathbf{f}^{\mathrm{ext}}_{n+1} - \mathbf{K} \mathbf{u}_{n+1} \right)$ \tcp*{\eqref{eq:nsn_smooth_equ}}
        }
        Advance \;
        $\mathbf u_n\gets \mathbf u_{n+1}$,
        $\mathbf v_n\gets \mathbf v_{n+1}$,
        $\mathbf a_n\gets \mathbf a_{n+1}$
    }
\end{algorithm}

\subsection{Modified \acrfull{TSL}}
\label{ss:modified_tsl}

\medskip\noindent
Unlike the formulation in \cite{collins2022formulation,collins2025formulation}, where the decohesion process is treated through a complementarity condition, similar to~\eqref{eq:decohesion_complementarity}, our model integrates decohesion explicitly by introducing the history variable $\delta_\text{max}$~\eqref{eq:camacho_ortiz_irreversibility}. This choice is motivated by a key observation in \cite{ghesquiere2025stability}: the discontinuity in the cohesive traction at softening events, i.e., when $\dot d > 0$, does not introduce any instability, as the energy error due to this jump is exactly compensated by the energy dissipated through decohesion. It allows us to preserve the high performance and simplicity of the explicit framework. However, while extrinsic cohesive zone models typically assume a rigid pre-fracture state, their implementation in an explicit dynamic framework introduces a critical numerical challenge during the onset of softening. As damage $d$ approaches zero, the secant stiffness $k(d)$ tends toward infinity~\eqref{eq:secant_stiffness}. In explicit schemes, where the stable time step $\Delta t$ is inversely proportional to the square root of the maximum system stiffness~\eqref{eq:stable_timestep}, this results in a vanishingly small $\Delta t$, rendering the simulation computationally infeasible. The contact implementation adopted here relies on the convexity of the underlying minimization problem~\eqref{eq:qp}. Because this convexity is lost when the stability condition is violated, and to avoid the need for impractically small time steps, we introduce a regularization of the traction-separation law by imposing a stiffness cap $\tilde{k}$. By setting the stiffness to zero (constant traction unloading) when $k > \tilde{k}$, we prevent the time step from dropping below a usable threshold, usually governed by the bulk material. Our capped law retains a secant unloading path whenever $k < \tilde{k}$.

\begin{figure*}[h!]
    \centering
    \includegraphics[scale=1]{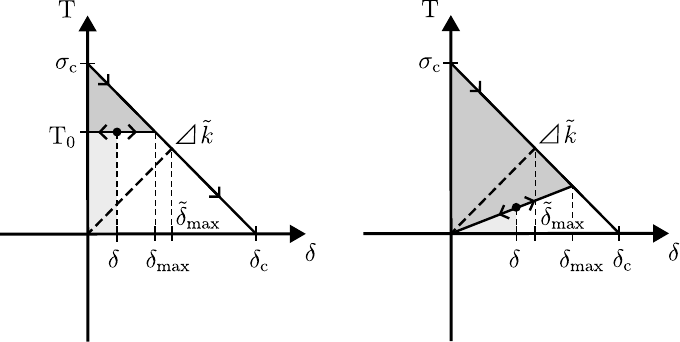}
    \caption{Modified traction-separation law with capped stiffness $\tilde{k}=k(\tilde d)$.}
    \label{fig:tsl_capped}
\end{figure*}

\medskip\noindent
To implement this regularized law, we link the globally defined stiffness threshold to a local damage threshold $\tilde{d}$ with $\tilde{k} = T(\tilde{d})/(\tilde{d}\delta_c)$. With the traction scalar function $T(d)$ defined as
\begin{equation}
    T(d) = \sigma_c (1-d),
\end{equation}
the damage threshold reads
\begin{equation}
    \tilde{d} = \frac{\sigma_c}{\sigma_c + \tilde{k} \delta_c}.
\end{equation}
Therefore, we split the law into two regimes and express the cohesive traction as
\begin{equation}
    \mathbf t^{\mathrm{coh}} 
    = \begin{cases}
        \sigma_c (1-d) \frac{\left(\delta_n \mathbf n + \beta^2 \boldsymbol{\delta}_t \right)}{\|\left(\delta_n \mathbf n + \beta^2 \boldsymbol{\delta}_t \right)\|}
        , & \text{if } d < \tilde{d}, \\
        \frac{1-d}{d} \frac{\sigma_c}{\delta_c} \frac{1}{\delta} \left(\delta_n \mathbf n + \beta^2 \boldsymbol{\delta}_t \right), & \text{if } d \ge \tilde{d}, 
    \end{cases}
\end{equation}
to which the irreversibility condition~\eqref{eq:camacho_ortiz_irreversibility} still applies. This modified law is illustrated in Figure~\ref{fig:tsl_capped}.

\medskip\noindent
The modified law affects the response of low-damage cohesive elements. The energy required to open at constant traction exceeds that required to open along a linear unloading/reloading path. We expect this effect to be negligible at high strain rates but potentially more significant at low strain rates. Indeed, a higher strain rate leads to greater initial damage in the cohesive elements, thereby causing overshooting of the regularized regime. The value of $\tilde{k}$ should be chosen high enough to minimize this effect. We can choose this value based on a fraction of the characteristic stiffness of a bulk element: $\tilde{k}=\alpha\frac{E}{h_e}$ with $h_e$ the characteristic size of a bulk element and $\alpha$ a user-defined parameter. For the simulations presented in this work using the nonsmooth contact implementation, we set $\alpha\in[1, 10]$ to minimize the effect of the modified law on the overall response while maintaining a practical time step. 

%% file: 03_numerical_validation.tex
\section{Numerical validation}
\label{sec:numerical_validation}

In this section, we validate the proposed semi-explicit \acrshort{NSN} scheme using three benchmark problems: the bouncing ball, the impacting bar, and the impacting bar with internal damage. The classical bouncing ball---a \acrfull{SDOF} system with rigid impact---serves as a fundamental test of the algorithm's accuracy in capturing impact and restitution phenomena. The second benchmark simulates the impact of a multi-degree-of-freedom bar against a rigid wall. For both benchmarks, the results are compared against the analytical solutions as well as two established nonsmooth contact dynamics schemes: the implicit \acrfull{MJ} scheme (with $\theta=\tfrac12$)~\cite{moreau1999numerical} and the semi-explicit \acrfull{CDL} scheme~\cite{fekak2017new}. We further evaluate the \acrshort{NSN} scheme against a traditional penalty-based contact formulation within the standard explicit Newmark-$\beta$ scheme (with $\beta=0$, $\gamma=\tfrac12$). For this purpose, and to model a case characteristic of dynamic fragmentation, the impacting bar problem is modified by introducing internal damage within the bar to represent a fragment that drifts after fragmentation and strikes a rigid obstacle.

\subsection{Bouncing ball}

\medskip\noindent
This problem focuses on the impact of a \acrshort{SDOF} system (a ball) on a rigid surface under the influence of gravity. The ball is modeled as a point mass $m$ with an initial height $u_0$ and zero initial velocity $v_0 = 0$. When the ball impacts the rigid surface at $u=0$, it undergoes an instantaneous change in velocity governed by the coefficient of restitution $e$, which defines the ratio of the post-impact relative velocity to the pre-impact relative velocity. The analytical solution for the ball's motion consists of a series of parabolic trajectories between impacts, where the time of each bounce and the maximum height reached are determined by the initial conditions and the coefficient of restitution. When $e<1$, the motion results in an infinite accumulation of impacts within a finite time, a phenomenon known as Zeno's paradox. In this section, we compare the displacement time histories obtained using the semi-explicit \acrshort{NSN} scheme with the analytical solution and the two alternative nonsmooth time-integration schemes, for both elastic ($e=1$) and inelastic ($e<1$) collisions. Although the ball is a scalar \acrshort{SDOF} system, we present the schemes in their general matrix–vector form to maintain consistency with the bar problem that follows.

\medskip\noindent
The \acrshort{MJ} scheme with $\theta=\tfrac12$ is an implicit time-integration method derived by integrating the distributional form of the equations of motion~\eqref{eq:distributional_motion} over the time step $[t_n, t_{n+1}]$. Displacements are updated using a midpoint rule ($\theta$-method with $\theta=\tfrac12$). Note that for any vector $\mathbf{z}$, the $\theta$-interpolated value is defined as
$\mathbf{z}_{n+\theta} = (1-\theta)\mathbf{z}_n + \theta\mathbf{z}_{n+1}$. The general case for any value of $\theta$ leads to the following set of equations:

\begin{subequations}
    \begin{empheq}[left=\empheqlbrace]{align}
        &\mathbf{M} \left(\mathbf{v}_{n+1} - \mathbf{v}_n\right) + \Delta t \, \mathbf{K}\,\mathbf{u}_{n+\theta} = \Delta t\,\mathbf{f}^{\mathrm{ext}}_{n+\theta} + \left(\mathbf{H}^\mathcal{A}\right)^\top \mathbf{p}^\mathcal{A}_{n+1}, \\
        & 0 \leq \mathbf{p}^\mathcal{A}_{n+1} \perp \mathbf{H}^\mathcal{A} \mathbf{v}_{n+1} + e \mathbf{H}^\mathcal{A}\mathbf{v}_{n} \geq 0, \quad \text{with } \mathcal{A} = \left\{ i \mid g_i\left(\mathbf{u}^{\text{pred}}_{n+1}\right) \le 0 \right\}\\
        & \mathbf{u}_{n+1} = \mathbf{u}_n + \Delta t \, \left((1-\theta) \mathbf{v}_n + \theta \mathbf{v}_{n+1}\right),
    \end{empheq}
\end{subequations}
with $\mathbf{u}^\text{pred}_{n+1} = \mathbf{u}_n + \Delta t \, \mathbf{v}_n$ the predicted displacement at step $n+1$. 

\medskip\noindent
While various explicit schemes for nonsmooth dynamics exist \cite{carpenter1991lagrange, paoli2002numericalI, paoli2002numericalII}, the \acrshort{CDL} scheme \cite{fekak2017new} was selected for its formulation within the \acrshort{NSCD} framework. In the same spirit as \acrshort{NSN}, the \acrshort{CDL} scheme treats contact implicitly within an explicit scheme; however, they differ fundamentally in how the contact impulse feeds back into the displacement and velocity. From a smooth-nonsmooth split of the dynamics, \acrshort{NSN} employs a predictor–corrector approach, that corrects both the velocity and the displacement within the same step, whereas \acrshort{CDL} adopts a staggered half-step discretization: the contact active set $\mathcal{A}$ is determined by the displacements at step $n+1$, which then governs the impulse $\mathbf{p}_{n+\frac{3}{2}}$ used to update velocities for the same subsequent half-step $n+\frac32$. In this configuration, the displacement $\mathbf{u}_{n+1}$ remains purely explicit and uncorrected by the current contact forces. Formally, the \acrshort{CDL} scheme is derived by integrating the distributional equations of motion~\eqref{eq:distributional_motion} over the interval $\left[t_{n+\frac12}, t_{n+\frac{3}{2}}\right]$:

\begin{subequations}
    \begin{empheq}[left=\empheqlbrace]{align}
        &\mathbf{M}\left(\mathbf{v}_{n+\frac{3}{2}} - \mathbf{v}_{n+\frac{1}{2}}\right) + \Delta t \, \mathbf{K}\,\mathbf{u}_{n+1} = \Delta t\,\mathbf{f}^{\mathrm{ext}}_{n+1} + \left(\mathbf{H}^\mathcal{A}\right)^\top \mathbf{p}^\mathcal{A}_{n+\frac{3}{2}}, \label{eq:cdl_v}\\
        & 0 \leq \mathbf{p}^\mathcal{A}_{n+\frac{3}{2}} \perp \mathbf{H}^\mathcal{A} \mathbf{v}_{n+\frac{3}{2}} + e \mathbf{H}^\mathcal{A} \mathbf{v}_{n+\frac{1}{2}} \geq 0, \quad \text{with } \mathcal{A} = \left\{ i \mid g_i\left(\mathbf{u}_{n+1}\right) \le 0 \right\}\\
        & \mathbf{u}_{n+1} = \mathbf{u}_n + \Delta t \, \mathbf{v}_{n+\frac{1}{2}} \label{eq:cdl_u}
    \end{empheq}
\end{subequations}
The \acrshort{CDL} scheme can be understood as an adaptation of the symplectic Euler integrator to the contact setting. Introducing the half-step velocity as a staggered state variable, i.e., performing the change of variables $(\mathbf{u}_n, \mathbf{v}_n) \mapsto (\mathbf{u}_n, \mathbf{v}_{n+\frac12})$, reveals an underlying partitioned structure: the displacement is advanced explicitly through $\mathbf{u}_{n+1} = \mathbf{u}_n + \Delta t\,\mathbf{v}_{n+\frac12}$, while the implicit coupling is confined entirely to the velocity and contact force updates at the subsequent half-step. This staggering is geometrically motivated: contact detection occurs at the integer instant $t_{n+1}$ via $\mathbf{u}_{n+1}$, and the nonsmooth velocity jump is localized naturally between the pre-impact velocity $\mathbf{v}_{n+\frac12}$ and the post-impact velocity $\mathbf{v}_{n+\frac32}$. Because the kinematic advance remains uncoupled from the current contact impulse, the active set is evaluated at the exact updated configuration, and the contact \acrshort{LCP} retains a particularly efficient, decoupled structure.

\begin{figure*}[h!]
    \centering
    \includegraphics[scale=1]{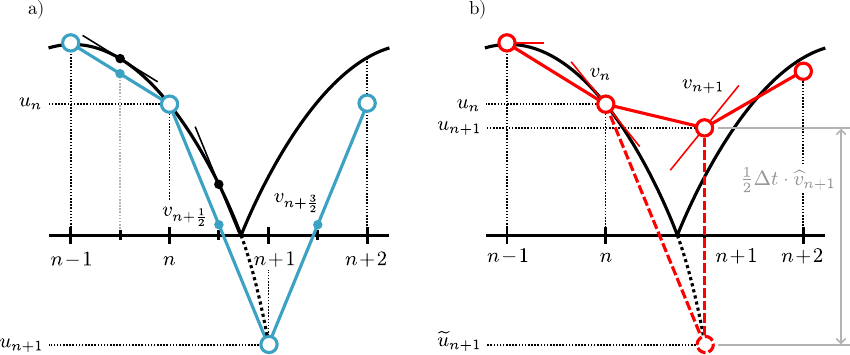}
    \caption{Schematic comparison of the time discretization and contact update logic of \acrshort{CDL} (left) and \acrshort{NSN} (right). In both \textbf{a)} and \textbf{b)}, the solid black line represents the analytical solution $u(t)$. The outlined circles represent the positions at full steps, and velocities are shown as slopes. In \textbf{a)}, the staggered structure of the \acrshort{CDL} scheme is presented, with small blue circles representing when the velocities are evaluated, at half-steps. The \acrshort{CDL} scheme detects contact at $t_{n+1}$ and corrects the velocity at $t_{n+\frac32}$, using the impact law that relates $v_{n+\frac32}$ to $v_{n+\frac12}$. \textbf{b)} The \acrshort{NSN} scheme employs a predictor-corrector approach. If the smooth displacement update $\widetilde{u}_{n+1}$ violates the contact constraint (dashed outlined circle), a velocity correction $\widehat{v}_{n+1}$ based on the impact law, is used to correct simultaneously the velocity $v_{n+1}$ and displacement $u_{n+1}$ at the same step contact is detected.}
    \label{fig:one_step_cdl_nsn}
\end{figure*}

\medskip\noindent
In contrast, the \acrshort{NSN} scheme employs a predictor-corrector approach. By using a smooth predictor $\widetilde{\mathbf{u}}_{n+1}$ to detect potential contact, the algorithm can solve for a corrective velocity $\mathbf{\widehat{v}}_{n+1}$. It enables simultaneous updates of velocities and displacements in a single step. While the \acrshort{NSN} corrector pulls $\mathbf{u}_{n+1}$ toward the feasible domain, small residual penetrations may persist because the correction is linearized at the velocity level. Conversely, the \acrshort{CDL} scheme does not correct $\mathbf{u}_{n+1}$: the integer-step displacement remains purely explicit, and any overlap with the obstacle is resolved only in the subsequent half-step through the velocity jump. The distinction between these two strategies---staggered in \acrshort{CDL} versus predictor-corrector in \acrshort{NSN}--- is illustrated for a bouncing ball problem in Figure~\ref{fig:one_step_cdl_nsn}.

\begin{figure*}[h!]
    \centering
    \includegraphics[scale=1]{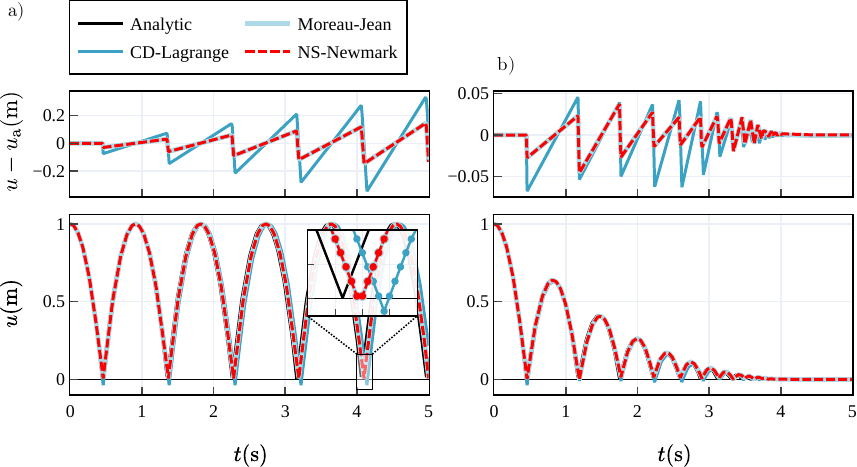}
    \caption{Comparison of displacement $u$ and distance to the analytical solution $u-u_{\text{a}}$ time histories for the implicit Moreau-Jean scheme, the semi-explicit CD-Lagrange scheme, and the semi-explicit \acrshort{NSN} scheme, for a coefficient of restitution of \textbf{a)} $e=1$ and \textbf{b)} $e=0.8$. Both \acrshort{NSN} and \acrshort{MJ} yield identical updates and a slightly better accuracy than the \acrshort{CDL} scheme.}
    \label{fig:u_v_comparison}
\end{figure*}

\medskip\noindent
To be consistent with the simulation parameters of Di Stasio et al.~\cite{di2019benchmark}, the mass of the ball is set to $m=1$~kg, the initial height to $u_0=1$~m, and the gravitational acceleration to $g=9.81$~m/s$^2$. Simulations are run for a total time of $T=5$~s, with a time step set initially to $\Delta t = 10^{-2}$~s. The displacement time histories, as well as their deviation from the analytical solution $u - u_{\text{a}}$, are presented in Figure~\ref{fig:u_v_comparison} for both elastic ($e=1$) and inelastic ($e=0.8$) collisions.

\medskip\noindent
All three schemes accurately capture the bouncing-ball dynamics, exhibiting only small shifts due to the lower-order accuracy at contact times. For both elastic and dissipative contacts, the \acrshort{NSN} scheme matches the results of the \acrshort{MJ} scheme exactly. This equivalence occurs when, under constant acceleration, the trapezoidal integration in \acrshort{MJ} becomes algebraically equivalent to the central-difference integration in \acrshort{NSN}---both during free flight and at the impact step, where acceleration terms cancel out. Provided that contact detection happens simultaneously, both schemes yield identical updates (see Appendix~\ref{appendix:bouncing_ball} for further details). Additionally, under these specific initial conditions and time-step choice, the \acrshort{CDL} scheme exhibits a slight phase shift relative to the other two schemes, as shown in the error plot and consistent with observations in \cite{di2019benchmark}. For these two test cases, the \acrshort{NSN} scheme achieves accuracy comparable to, or better than, that of the two established nonsmooth contact dynamics schemes studied here.

\medskip\noindent
For a more systematic study, we examine the effect of time-step refinement on the accuracy of the schemes by performing a convergence study for both $e=1$ and $e=0.8$. Approximately 50 time step values ranging from $\Delta t = 10^{-6}$~s to $\Delta t = 10^{-1}$~s on a log scale are considered. The error in displacement $\eta$ is computed using the following normalized $L^1$ norm \cite{acary2012higher,fekak2017new,di2019benchmark}:

\begin{equation}
    \eta = \frac{\sum_{i=1}^N \|u_i - u_\text{a}(t_i)\|}{\sum_{i=1}^N\|u_\text{a}(t_i)\|},
    \label{eq:displacement_error}
\end{equation}

\begin{figure}[h!]
    \centering
    \includegraphics[scale=1]{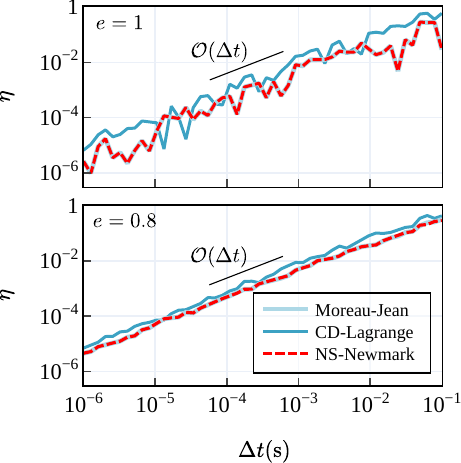}
    \caption{Comparison of the convergence rate of the error in displacement $\eta$ for the implicit Moreau-Jean scheme, the semi-explicit \acrshort{CDL} scheme, and the semi-explicit \acrshort{NSN} scheme, for coefficients of restitution of $e=1.0$ and $e=0.8$. Both the \acrshort{MJ} and \acrshort{NSN} schemes achieve similar accuracy, with slightly lower errors than the \acrshort{CDL} scheme across the range of time steps considered.}
    \label{fig:u_convergence}
\end{figure}

\medskip\noindent
where $u_i$ is the numerical displacement at step $i$, $u_\text{a}(t_i)$ is the analytical displacement at time $t_i$, and $N$ is the total number of time steps in the simulation. Figure~\ref{fig:u_convergence} shows the convergence of the displacement error $\eta$ with respect to the time step $\Delta t$ for the three schemes and two restitution coefficients. For $e=1$ and $e=0.8$, the schemes exhibit a convergence rate of $\eta \propto \mathcal{O}(\Delta t)$. This first-order convergence is consistent with previous studies on nonsmooth contact dynamics schemes and is explained by the first-order accuracy of impact events. For that specific test case where gravity is constant, both the \acrshort{MJ} and \acrshort{NSN} schemes achieve similar accuracy, with slightly lower errors than the \acrshort{CDL} scheme across the range of time steps considered. 

\subsection{Impacting bar}
\label{sec:impacting_bar}

While the previous rigid-body simulation serves as a fundamental test of the algorithm's ability to capture impact and restitution phenomena, it does not fully exploit the second-order accuracy of the two semi-explicit schemes during free-flight phases. We therefore consider a second benchmark problem, the impacting bar, which simulates the impact of a multi-degree-of-freedom system on a rigid wall. We adopt the benchmark parameters from~\cite{di2019benchmark,fekak2017new,carpenter1991lagrange}: a bar of length $L=0.254$~m, cross-sectional $A=6.45\cdot10^{-4}$~m$^2$, Young's modulus $E=211$~GPa and density $\rho=7847$~kg/m$^3$. The bar is launched with an initial velocity $v_0=5$~m/s, with contact occurring at a single node whose position and velocity are denoted by $u_c$ and $v_c$. 

\begin{figure}[h!]
    \centering
    \includegraphics[scale=1]{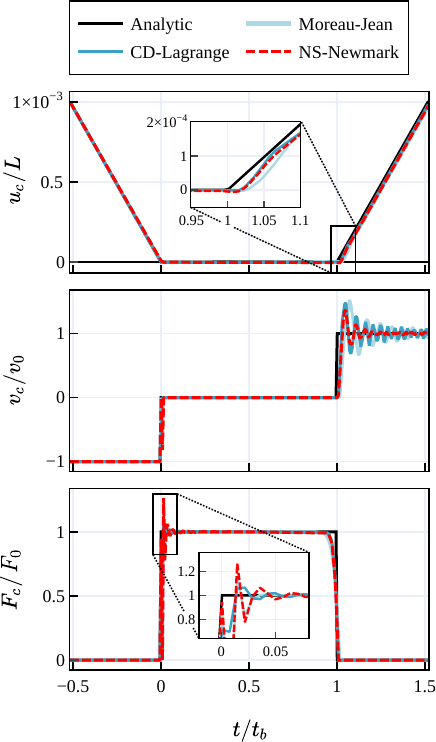}
    \caption{Comparison of semi-explicit \acrshort{NSN}, semi-explicit \acrshort{CDL} and implicit \acrshort{MJ} schemes with the 1D elastic bar impact analytical solution. The results show the normalized contact node position $(u_c/L)$ (top), velocity $(v_c/v_0)$ (middle), and contact force $(F_c/F_0)$ (bottom) over the normalized bouncing time $t/t_b$. All simulations use $N=50$ elements and a restitution coefficient of $e=0$.}
    \label{fig:bar_u_v_p_comparison}
\end{figure}

\medskip\noindent
We discretize the bar into $N_e=50$ P1 elements of uniform size with a time step of $\Delta t = 0.7 \times \Delta t_c \approx 6.874\cdot 10^{-7}$~s, with $\Delta t_c$ the \acrfull{CFL} condition. To better align with the analytical solution and eliminate non-physical oscillations immediately following the first impact, and to match results of~\cite{di2019benchmark}, the coefficient of restitution is set to $e=0$. This choice results in local dissipation of kinetic energy at the contact node. Results are validated against the analytical solution, which predicts a bouncing time of $t_b = 2 \times L / c$, where $c = \sqrt{E/\rho}$ is the wave speed of the bar. In the subsequent plots, we normalize position, velocity, and contact force by $L$, $v_0$, and $F_0 =\rho\cdot c \cdot v_0 \cdot A$, respectively, while time is normalized by the bouncing time $t_b$. Results are shown in Figure~\ref{fig:bar_u_v_p_comparison}.

\medskip\noindent
At the chosen time step, all three schemes accurately capture the system physics, closely matching the analytical bouncing time and the characteristic rectangular contact-force profile. However, a closer inspection of the transition phases reveals subtle numerical distinctions:

\begin{itemize}
    \item At the initial impact ($t/t_b=0$), the contact force shows slightly larger oscillations for the \acrshort{NSN} scheme compared to the others. These oscillations tend to be rapidly damped out.
    \item At the contact release ($t/t_b=1$), the inset in the position plot reveals a close match between the two second-order schemes (\acrshort{NSN} and \acrshort{CDL}), with a slight advantage for \acrshort{CDL}, while the \acrshort{MJ} scheme shows a slight shift. Additionally, at this time, the \acrshort{NSN} scheme exhibits the lowest residual oscillations in the velocity.
\end{itemize}

\begin{figure*}[h!]
    \centering
    \includegraphics[scale=1]{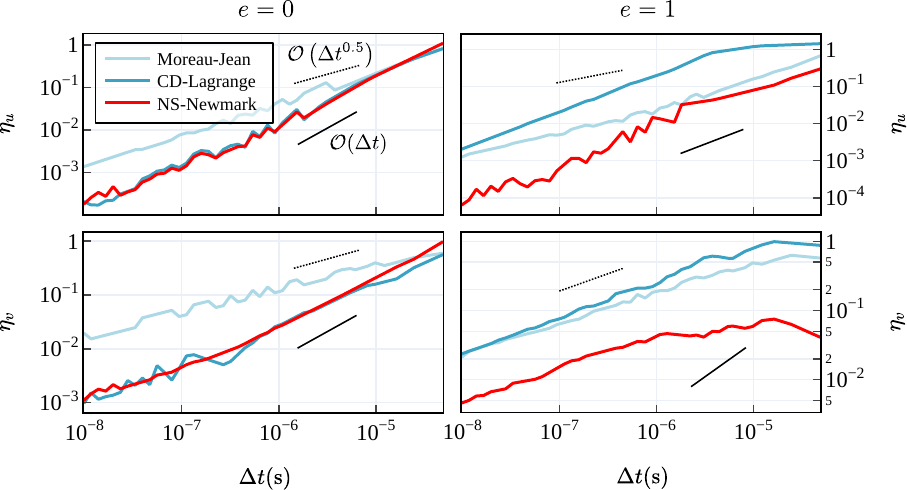}
    \caption{Time-step convergence of displacement errors $\eta_u$ (top) and velocity errors $\eta_v$ (bottom) for the impacting bar benchmark, comparing perfectly plastic $e=0$ (left) and elastic $e=1$ (right) impacts. Errors are evaluated from $t/t_b > 1$ for a duration of $t=3 t_b$, with coupled spatio-temporal refinement ($N_e\in [1, 5\cdot10^3]$, $\Delta t = 0.999 \Delta t_c$). The \acrshort{MJ} scheme exhibits $\mathcal{O}(\Delta t^{0.5})$ convergence in all cases. Both \acrshort{NSN} and \acrshort{CDL} achieve $\mathcal{O}(\Delta t)$ convergence, except for velocity at $e=1$ where the rate drops to $\mathcal{O}(\Delta t^{0.5})$. Notably, \acrshort{NSN} yields significantly lower error magnitudes than \acrshort{CDL} in the purely elastic regime.}
    \label{fig:bar_u_v_convergence}
\end{figure*}

\medskip\noindent
To further investigate the accuracy of the schemes, we perform a convergence study by refining the mesh and the time step simultaneously ($N_e\in [1, 5\cdot10^3]$, $\Delta t = 0.999 \Delta t_c$). The $L^1$ errors \eqref{eq:displacement_error} for displacement, $\eta_u$, and velocity, $\eta_v$, are computed with respect to the analytical solution after contact release ($t/t_b > 1$) for a duration of $t=3 t_b$. While $e=0$ provides a clean comparison by dissipating energy at the contact node, we also evaluate the perfectly elastic limit ($e=1$). Testing $e=1$ allows us to observe how the schemes handle frequent discrete velocity jumps at the contact interface in the absence of contact dissipation. By examining both the $e=0$ and $e=1$ cases, we verify the robustness of the convergence rates across the full range of impact physics. The results are shown in Figure~\ref{fig:bar_u_v_convergence}.

\medskip\noindent
The convergence results indicate that non-constant acceleration significantly alters the performance of the numerical schemes, favoring second-order schemes in free flight. For the perfectly plastic case ($e=0$), both \acrshort{NSN} and \acrshort{CDL} maintain a robust convergence rate of $\mathcal{O}(\Delta t)$ for both displacement and velocity. Their error magnitudes are nearly identical, yielding a significant efficiency gain over the \acrshort{MJ} scheme, which degrades at $\mathcal{O}(\Delta t^{0.5})$ across all tested configurations. Evaluating the perfectly elastic limit ($e=1$) reveals critical differences. Without dissipation at the contact node to stabilize the interface, the velocity convergence rate ($\eta_v$) for both \acrshort{NSN} and \acrshort{CDL} degrades to $\mathcal{O}(\Delta t^{0.5})$. However, they successfully maintain $\mathcal{O}(\Delta t)$ for displacement ($\eta_u$). More importantly, while their asymptotic rates are similar in this limit, the absolute error magnitudes diverge significantly: the \acrshort{NSN} scheme yields substantially lower errors than \acrshort{CDL} across the tested time steps. This result underscores the advantage of the simultaneous predictor-corrector approach used in \acrshort{NSN}, which proves more robust than the staggered update logic of \acrshort{CDL} when handling the repeated velocity jumps at the contact interface.

\medskip\noindent
Taken together, these benchmark results indicate that the \acrshort{NSN} scheme is a reliable formulation capable of handling a broad spectrum of contact dynamics, from isolated rigid-body collisions to the sustained contact interactions found in elastic systems. Having established its theoretical and practical standing among nonsmooth integrators, we next evaluates the semi-explicit \acrshort{NSN} scheme in a scenario representative of dynamic fragmentation. It compares its performance with traditional penalty-based formulations for contact.

\subsection{Impacting bar with internal damage}
\label{ss:impacting_bar_with_internal_damage}

\medskip\noindent
In fragmentation scenarios, where interactions between crack faces and multiple fragments are omnipresent, traditional penalty-based formulations are preferred for their computational efficiency. In these methods, penetration is resisted by a force proportional to the penalty stiffness $\epsilon_n$ and to the penetration magnitude. However, selecting an appropriate value for $\epsilon_n$ necessitates a compromise between accuracy and stability. While high stiffness is required to minimize unphysical penetration, it also limits the stability to prohibitively small time steps. Additionally, as demonstrated in \cite{ghesquiere2025stability}, the nonsmooth loading path induced by this artificial contact stiffness generates energy artifacts in explicit time-integration schemes. If an interface oscillates around a null gap, repeatedly entering and leaving contact, artifacts often accumulate, potentially leading to global instability in the simulation. We present the limitations of such an approach in the following study and compare it with the \acrshort{NSN} scheme. To that end, we adapt the previous benchmark to simulate the impact of a bar with internal damage. The internal damage is modeled using cohesive elements. Theoretically, the damage affects only the material's 1D tensile behavior and should therefore not affect its compressive behavior. It is expected that the bouncing time $t_b$ is the same as the non-damaged scenario. However, when using the penalty method for contact, we expect the contact penalty to affect the simulation's accuracy and stability, as they influence the bar's compliance under compression.

\medskip\noindent
We remark that while standard \acrfull{CFL} conditions ($\Delta t \le h_\text{min}/c$) govern the bulk stability, they fail to account for the additional stiffness $k_\text{add}$ from cohesive zones or contact penalties. These high-stiffness terms increase the system's maximum natural frequency, $\omega_{\max}$, thereby reducing the stable time step. To ensure the numerical stability of each case, we calculate the critical time step using a conservative estimate based on Gershgorin's circle theorem~\cite{osti_5429002}. Using a lumped mass matrix $\mathbf{M}$, Gershgorin's theorem bounds the maximum eigenvalue of $\mathbf{M}^{-1}\mathbf{K}$, $\lambda_{\max} =\omega_{\max}^2$ by the maximum row sum of the stiffness matrix scaled by the nodal mass:
\begin{equation}
    \omega_\text{max}^2 \le \max_i\left(\frac{\sum_j|\mathbf{K}_{ij}^\text{bulk}|}{\mathbf{M}_{ii}} + \frac{k_\text{add}}{\mathbf{M}_{ii}}\right) \approx \omega_\text{bulk}^2 + \omega_\text{add}^2
\end{equation}
We compute the stable time step using the updated $\omega_{\max}$, $\Delta t \le 2/\omega_{\max}$, thereby ensuring theoretical stability. In practice, additional safety factors are necessary to account for nonsmooth loading paths, further reducing the permissible $\Delta t$. Note that in one dimension and with $k_\text{add} = 0$, this bound gives exactly the standard \acrshort{CFL} condition. 

\medskip\noindent
To fit the subsequent simulations, the bar has a length of $L=10^{-3}$~m and is made of AD-995 alumina, with material properties listed in Table~\ref{tbl:material}. It is discretized with $N_e=2\cdot 10^3$ P1 elements. To model high levels of internal damage, $N_\text{coh}=10^3$ cohesive elements, whose damage is initialized at $d_0=10^{-3}$, are inserted into the mesh. This damage level yields a cohesive stiffness $k(d_0)$ (Eq.~\ref{eq:secant_stiffness}) smaller than the characteristic stiffness of a bulk element $E/h_e$, with $h_e=L/N_e$, ensuring the convexity of the contact problem is not influenced by cohesion. For the penalty-based comparison, we define a contact penalty $\epsilon_n = \alpha E/h_e$ with $\alpha\in[10^{-2}, 10^2]$. 

\begin{table}[h!]
  \centering
  \begin{tabular}{l c}
    \toprule
    \textbf{Parameter} & \textbf{Value} \\
    \midrule
    Young’s modulus, $E$ [GPa]                      & 370              \\
    Density, $\rho$ [kg/m\(^3\)]                    & 3900             \\
    Cohesive strength $\sigma_c$ [MPa]              & 262              \\
    Fracture toughness, $G_c$ [J/m\(^2\)]           & 50               \\
    \bottomrule
  \end{tabular}
  \caption{AD‑995 alumina material parameters.}
  \label{tbl:material}
\end{table}

\begin{figure*}[h!]
    \centering
    \includegraphics[scale=1]{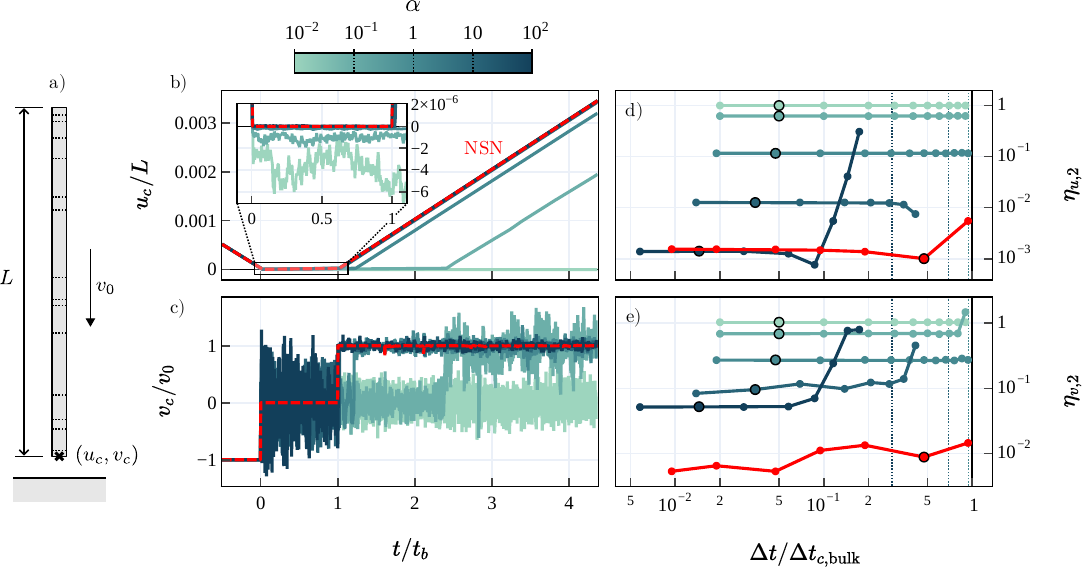}
    \caption{\textbf{(a)} Schematic of a bar of length $L$ and material properties $E, \rho$, discretized with $N=2\cdot10^3$ P1 elements, with $N_\text{coh}=10^3$ inserted cohesive elements (dashed lines) of properties $\sigma_c, G_c, d$, to represent a highly damaged bar, launched against a rigid obstacle with a rigid-body velocity of $v_0$. Comparison of the contact node position $u_c$ \textbf{(b)} and velocity $v_c$ \textbf{(c)} with respect to time $t/t_b$ for the penalty-based method with different normalized penalty values $\hat{\epsilon_n} = \epsilon_n / (E/\bar{h}_e)$, compared to the nonsmooth solution (red line). Figure \textbf{(d)} and \textbf{(e)} show the $L^2$ norm of the error in the contact node position and velocity against the analytical solution after impact,  as a function of the normalized time step $\Delta t / \Delta t_{c,\text{bulk}}$. $\Delta t_{c, \text{bulk}}$ represents the stable time step of the bulk elements without the additional stiffness contribution of contact or cohesion. The corresponding stable time steps, accounting for additional stiffness, are shown as dashed lines. The red line indicates the error for the nonsmooth solution. Specific time evolution of displacement and velocity from \textbf{(b)-(c)} is indicated with black outlined markers on \textbf{(d)-(e)}.}
    \label{fig:bouncing_bar_convergence}
\end{figure*}

\medskip\noindent
\paragraph{Displacement and velocity time evolution and convergence.}
Figure~\ref{fig:bouncing_bar_convergence} shows the results of the bouncing bar (with internal damage), in terms of displacement and velocity of the contact node $(u_c, v_c)$, and the corresponding $L^2$ errors. The specific displacements and velocity of Fig.~\ref{fig:bouncing_bar_convergence}b-c are indicated on Fig.~\ref{fig:bouncing_bar_convergence}d-e as black outlined markers. As evidenced in Fig.~\ref{fig:bouncing_bar_convergence}b, the penalty method exhibits strong sensitivity to the normalized contact stiffness, $\alpha$. Low stiffness values lead to significant unphysical penetration and delayed rebound, thereby failing to capture the kinematics accurately. Conversely, increasing the stiffness to enforce non-penetration better introduces high-frequency oscillations in the velocity field (Fig.~\ref{fig:bouncing_bar_convergence}c). It also severely restricts the stability region (Fig.~\ref{fig:bouncing_bar_convergence}d-e, vertical dashed lines). For an acceptable error on position, the penalty method ideally requires $\alpha \ge 10^2$, which is stable from $\Delta t /\Delta t_{c,\text{bulk}} \approx 5\cdot 10^{-2}$. In contrast, the nonsmooth solution (red line) enforces the contact condition without tuning a penalty parameter, maintaining low, stable errors in both position and velocity over a wide range of time steps. It achieves displacement accuracy comparable to the penalty method at $\alpha \approx 10^2$ and significantly lower velocity errors, even with $\Delta t /\Delta t_{c,\text{bulk}}\approx 1 $.

\begin{figure*}[h!]
    \centering
    \includegraphics[scale=1]{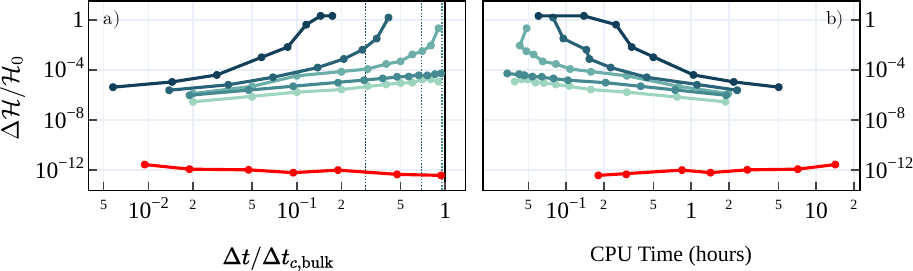}
    \caption{\textbf{(a)} Convergence analysis of the relative algorithmic energy error $\Delta\mathcal{H}/\mathcal{H}_0$ as a function of the normalized time step $\Delta t / \Delta t_{c,\text{bulk}}$. The nonsmooth solution (red line) maintains energy conservation at machine-precision levels ($\approx 10^{-12}$), effectively independent of the time-step size. In contrast, the standard penalty-based schemes (blue lines) exhibit a characteristic convergence order, with accuracy degrading as $\Delta t$ increases, eventually diverging near their respective stability limits (vertical dashed lines). \textbf{(b)} Computational efficiency assessment plotting the energy error against the total CPU time. Although the semi-implicit nature of the nonsmooth formulation incurs a computational cost approximately 5 times higher per time step, its unconditional energy stability allows the use of significantly larger time steps, resulting in a lower total time-to-solution for high-precision results compared to the penalty-based method.}
    \label{fig:ealgo_comparison}
\end{figure*}

\medskip\noindent
\paragraph{Energy conservation and computational cost.}While the explicit Newmark-$\beta$ scheme only conserves the total mechanical energy $\mathcal{E}$ in a weak sense (oscillating around the mean), it conserves the algorithmic energy $\mathcal{H}$ exactly for a smooth evolution of the system. This quantity augments the mechanical energy with an acceleration-dependent term~\cite{hughes1977note,acary2016energy}:
\begin{equation}
    \mathcal{H} = \underbrace{\frac{1}{2} \mathbf{v}^\top \mathbf{M} \mathbf{v} + \frac{1}{2} \mathbf{u}^\top \mathbf{K} \mathbf{u}}_{\mathcal{E}} - \frac{\Delta t^2}{8} \mathbf{a}^\top \mathbf{M} \mathbf{a}
    \label{eq:algorithmic_energy}
\end{equation}
As proven in \cite{acary2016energy}, \acrshort{NSN} guarantees this conservation in the presence of contact as well, provided they are non-dissipative. Otherwise, dissipation is controlled by the coefficient of restitution $e$. This theoretical property is numerically verified in Figure~\ref{fig:ealgo_comparison}. The left panel plots the maximum relative error $\Delta\mathcal{H} / \mathcal{H}_0$ against the normalized time step. The \acrshort{NSN} method (red line) exhibits a flat error profile hovering around $10^{-12}$, independent of the time step. This value can be attributed to the accumulation of errors of the order of $10^{-14}$, the set tolerance of the contact solver. In contrast, the penalty-based method exhibits energy errors that converge to a plateau between $10^{-5}$ and $10^{-7}$ for small time steps, with lower accuracy for larger penalty values. The right panel of Figure~\ref{fig:ealgo_comparison} translates this into a practical efficiency metric: energy error versus total CPU time. All the simulations were run on an Intel(R) Xeon(R) Platinum 8360Y CPU @ 2.40 GHz. 
While the implicit nature of the \acrshort{NSN} contact solver increases per-step costs by a factor of $\sim5$, its robustness permits significantly larger time steps, leading to a faster total time-to-solution. This efficiency is achieved without compromising accuracy; in fact, \acrshort{NSN} yields an energy-conservation error nearly seven orders of magnitude lower than that of the stiffest penalty-based approach tested, while achieving a CPU time roughly 27 times smaller. By decoupling physical accuracy from the time-step constraints inherent to penalty methods, the nonsmooth approach effectively bypasses the traditional trade-off between contact stiffness and computational cost, enabling high-fidelity long-term simulations.

%% file: 04_application_to_dynamic_fragmentation.tex
\section{Application to dynamic fragmentation}
\label{sec:application_to_dynamic_fragmentation}

In this section, we apply the proposed contact algorithm to simulate the dynamic fragmentation of an expanding ring. This benchmark problem is widely used to validate fracture and fragmentation models because of its well-characterized behavior at high strain rates. We first present the standard free expansion ring fragmentation, and then detail a modified version that introduces walls to confine the process and increase contact interactions. To the best of our knowledge, fragment statistics in an expanding-ring configuration with rigid walls that confine the expansion and intensify fragment–fragment and fragment–wall interactions have not been systematically examined in the literature. This modified setup, therefore, provides a contact‑rich benchmark for assessing the robustness of the proposed contact algorithm and yields interesting physics.

\subsection{Free expanding ring fragmentation}
\label{ss:expanding_ring_fragmentation}
The Mott ring is a circular ring of radius $r$ subjected to rapid uniform radial expansion, driven by high internal pressure $p_\text{int}$, which induces high strain rates and eventual fragmentation due to tensile failure (Figure~\ref{fig:expanding_ring_equiv}a). While the failure is driven by tension, contact mechanics play a critical role in the post-fracture regime. Once fragmentation begins, the sudden release of stress generates compressive waves that trigger contact between the newly created cohesive surfaces. Secondary collisions between fragments may also occur. To capture this process numerically, we employ the extrinsic cohesive zone framework implemented in the open-source software \texttt{Akantu}~\cite{richart2024akantu}, coupled with a custom Python implementation of the \acrshort{NSN} scheme. 

\subsubsection{Problem setup}
\begin{figure}[h!]
    \centering
    \includegraphics[scale=1]{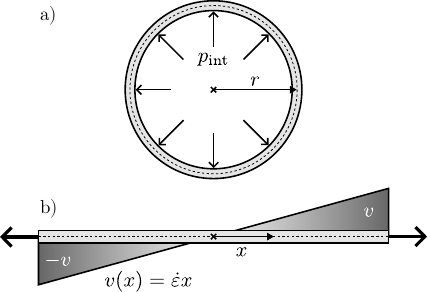}
    \caption{Schematic of the expanding ring fragmentation benchmark. The circular ring of radius $r$ under strong internal pressure $p_\text{int}$ ($\textbf{a}$) is modeled by a one-dimensional equivalent bar centered at $x=0$ ($\textbf{b}$) with an initial velocity field $v(x) = \dot{\varepsilon} x$ to simulate the expansion. The boundaries of the bar are pulled at a velocity $\pm\dot{\varepsilon} L/2$ until fragmentation starts.}
    \label{fig:expanding_ring_equiv}
\end{figure}

We study the one-dimensional equivalent of this system: a bar with length $L$, centered at the origin along the $x$-axis. To simulate the expansion of the ring, the bar is initialized with a velocity field $v(x) = \dot{\varepsilon} x$, where $\dot{\varepsilon}$ is the imposed constant strain rate (Figure~\ref{fig:expanding_ring_equiv}b). The boundaries of the bar are displaced with a velocity of $\pm\dot{\varepsilon} L/2$ to ensure uniform expansion, then released once fragmentation occurs. The material is modeled using the AD-995 alumina parameters listed in Table~\ref{tbl:material}.

\medskip\noindent
Following Camacho and Ortiz (1996), and Drugan (2001)~\citep{camacho1996computational, drugan2001dynamic}, we define the characteristic scales of the problem based on the material's elastic properties ($E$, $\rho$) and cohesive law properties ($\sigma_c$, $G_c$). First, the longitudinal wave speed is defined as $c = \sqrt{E/\rho}$. From the energy per unit area required to reach full decohesion (fracture energy $G_c$), we derive the characteristic material time $t_0$:
\begin{equation}
    t_0 = \frac{EG_c}{\sigma_c^2 c}.
\end{equation}
This timescale physically represents the duration of the cohesive fracture process, i.e., the time required for a cohesive interface to evolve from intact to fully separated under stresses on the order of $\sigma_c$ transmitted by longitudinal waves. The characteristic material length $s_0 = c t_0$ represents the distance a stress wave travels during the total decohesion process \citep{zhou2006effects}. Finally, we introduce the characteristic strain rate $\dot{\varepsilon}_0$ as the strain rate at which the elastic strain reaches the fracture deformation threshold ($\sigma_c/E$) over the characteristic time $t_0$:
\begin{equation}
    \dot{\varepsilon}_0 = \frac{\sigma_c}{E t_0}.
\end{equation}

\medskip\noindent
These scales allow us to express the imposed strain rate, the resulting fragment sizes, and the fracture energy in non-dimensional forms: $\hat{\dot{\varepsilon}}=\dot{\varepsilon}/\dot{\varepsilon}_0$,  $\hat{s} = s/s_0$, and $\hat{\mathcal{G}} = \mathcal{G}/G_c$ respectively. The transition from quasi-static to dynamic regime occurs at $\hat{\dot{\varepsilon}} \approx 1$.

\subsubsection{Energy budget}
\label{sssec:energy_budget_and_fragmentation_physics}
In this dynamic regime described by Grady’s model \citep{grady1982local}, fragmentation is governed by a local energy budget. As the bar of length $L$ expands at a constant strain rate $\dot{\varepsilon}$, the average fragment size $s$ emerges from a balance between the local kinetic energy and the fracture energy $G_c$. For a single fragment $i$ of size $s$, the local kinetic energy density is calculated by integrating the energy density over the fragment domain $[-s/2, s/2]$:
\begin{equation}
        \mathcal{K}_{\text{local},i} = \int_{-s/2}^{s/2} \frac{1}{2} \rho (\dot{\varepsilon} x)^2 dx = \frac{\rho \dot{\varepsilon}^2 s^3}{24}.
        \label{eq:local_kinetic_per_fragment}
\end{equation}
A fragment is formed when this available local kinetic energy is sufficient to overcome the material's resistance to fracture, defined by the fracture energy $G_c$. By equating the source ($\mathcal{K}_{\text{local},i}$) to the sink ($G_c$), the characteristic fragment size for free expansion is given by:
\begin{equation}
    \bar{s}_\text{free} = \left(\frac{24 G_c}{\rho \dot{\varepsilon}^2}\right)^{1/3},
    \label{eq:fragment_size_free}
\end{equation}
with a fragment size scaling with the strain rate as $s\propto\dot\varepsilon^{-2/3}$. In this regime, Grady assumes that only kinetic energy is available to drive fragmentation and that all fragments have the same size. For lower strain rates, the elastic energy stored in the bar makes a significant contribution to the energy budget. Glenn and Chudnovsky \citep{glenn1986strain} extended Grady's model to account for this energy contribution, which becomes dominant when ($\hat{\dot{\varepsilon}} < 1$), resulting in a model accounting for two distinct regimes: quasi-static and dynamic. Later, Zhou et al. \citep{zhou2006effects} proposed a model fitted to numerical results from a cohesive zone model that integrates the effects of wave propagation, contact, and damage evolution. The three normalized fragment size predictions from these models are given by:
\begin{equation}
    \hat{s}_\text{Grady} = \frac{\bar{s}_\text{free}}{s_0} = \left(\frac{24}{\hat{\dot{\varepsilon}}^2}\right)^{1/3}, \qquad
    \hat{s}_{\text{G\&C}} = \frac{4}{\hat{\dot{\varepsilon}}} \sinh \left[ \frac{1}{3} \sinh^{-1} \left( \frac{3}{2} \hat{\dot{\varepsilon}} \right) \right], \qquad
    \hat{s}_\text{ZMR} = \frac{4.5}{1 + 4.5\hat{\dot{\varepsilon}}^{2/3}}.
\end{equation}
ZMR model deviates from G. \& C. at low strain rates by predicting larger fragments. Although Glenn and Chudnovsky's model assumes that all the elastic energy is converted into fracture energy, the ZMR model accounts for the fact that the fragmentation is not instantaneous and that almost half of the elastic energy is kept as residual kinetic energy in the form of wave propagation, i.e., vibration of the fragments. At higher strain rates, the ZMR model predicts the same $s\propto \dot\varepsilon^{-2/3}$ scaling but with smaller fragments than both Grady and G. \& C. models, in closer agreement with experimental observations. In that regime, the system extracts more kinetic energy from the bar's global motion, notably due to contact interactions. To verify the ability of the proposed \acrshort{NSN} scheme to capture these contact interactions and their influence on fragmentation, we perform simulations for a wide range of such non-dimensional strain rates, specifically $\hat{\dot{\varepsilon}} \in [10^{-3}, 10^3]$, covering the quasi-static and dynamic range.

\subsubsection{Numerical implementation}

We conduct simulations of the expanding ring benchmark using both the penalty-based contact method and the proposed \acrshort{NSN} scheme. The mesh choices are detailed in Appendix~\ref{appendix:expanding_ring_simulation_parameters}, following the recommendations of ~\citep{molinari2007cohesive,zhou2005cohesive,levy2010dynamic} for converged numerical results. For the penalty-based approach, we select a penalty stiffness of $\epsilon_n = 10 E/\bar{h}_e$, with $\bar{h}_e$ the mean element length of the mesh, and the regular Camacho-Ortiz \acrshort{TSL} with no capping. \acrshort{NSN} uses a coefficient of restitution $e=1$, representing perfectly elastic collisions, and the modified \acrshort{TSL} introduced in Section~\ref{ss:modified_tsl}. The stiffness cap for cohesive elements is set to $\tilde{k}= 10 E/\bar{h}_e$, such that the two contact approaches share the same maximum stiffness, one for contact and the other for cohesion, thereby defining the same maximum stable time step $\Delta t_c$ (Section~\ref{ss:impacting_bar_with_internal_damage}). The actual time step is set to $\Delta t = 0.2 \Delta t_c$ for the penalty-based method to ensure stability and $\Delta t = 0.99 \Delta t_c$ for the nonsmooth method. The results are compared against the predictions from Grady \citep{grady1982local}, Glenn and Chudnovsky (G. \& C.) \citep{glenn1986strain}, and Zhou et al. (ZMR) \citep{zhou2006effects}. 

\medskip\noindent
While the first two models (Grady and G. \& C.) provide theoretical predictions for both fragment size and fracture energy, the ZMR model provides a closed-form prediction for fragment size only, fitted on numerical simulations. We approximate the fracture energy for the ZMR model as $\hat{\mathcal{G}}_{\text{ZMR}} = 1/\hat{s}_{\text{ZMR}}$, which represents the theoretical surface energy required to create fragments of size $\hat{s}_{\text{ZMR}}$. However, this approximation assumes that all energy is dissipated through fracture, neglecting the internal damage mechanisms present in the \acrshort{CZM} simulations of~\citep{zhou2006effects}. Therefore, $\hat{\mathcal{G}}_{\text{ZMR}}$ is shown as a theoretical lower bound.

\subsubsection{Results and discussion}
\begin{figure}[h!]
    \centering
    \includegraphics[scale=1]{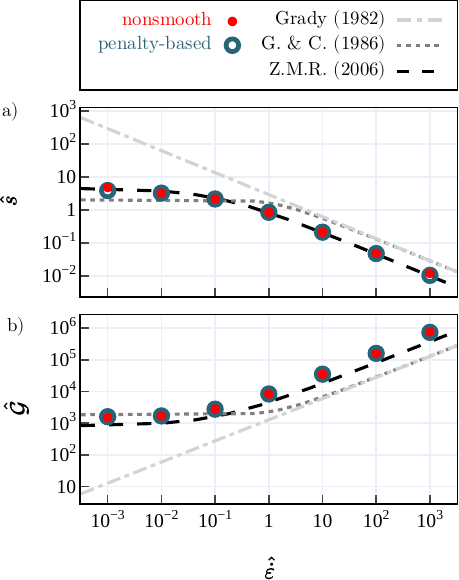}
     \caption{Comparison of fragmentation metrics for the expanding ring test. \textbf{a)} Average non-dimensional fragment size $\hat{s}$ as a function of the imposed non-dimensional strain rate $\hat{\dot{\varepsilon}}$. \textbf{b)} Non-dimensional fracture energy $\hat{\mathcal{G}}$ as a function of the imposed non-dimensional strain rate $\hat{\dot{\varepsilon}}$. The results for both penalty-based and nonsmooth contact methods are compared against theoretical predictions from Grady \citep{grady1982local}, Glenn and Chudnovsky \citep{glenn1986strain}, and the ZMR model \citep{zhou2006effects}. Both methods yield similar fragmentation behavior across the range of strain rates, with fragment sizes closely aligning with the ZMR predictions. A slight deviation at the lowest strain rate of the nonsmooth method is attributed to its modified traction-separation law. The fracture energy results indicate that both methods dissipate more energy than the ZMR lower bound, indicating the presence of cohesive elements with damage levels smaller than 1.}
    \label{fig:expanding_ring_results}
\end{figure}

Figure~\ref{fig:expanding_ring_results} presents both the average fragment size $\hat{s}$ and the fracture energy $\hat{\mathcal{G}}$ as a function of the imposed strain rate $\hat{\dot{\varepsilon}}$. For a consistent comparison, the dissipated energy was normalized by the bar length, such that we plot the fracture energy per unit length. We see in Figure~\ref{fig:expanding_ring_results}a that both contact methods yield fragment sizes that closely align with the ZMR predictions across the entire range of strain rates. At the lowest strain rate $\hat{\dot{\varepsilon}}=10^{-3}$, we observe a slight deviation: the nonsmooth method yields larger fragment sizes than the penalty-based approach. This discrepancy stems from the modified \acrshort{TSL} used in the \acrshort{NSN} scheme. As discussed in Section~\ref{ss:modified_tsl}, cohesive elements with low damage levels operate within the constant-traction regime of the modified law, thereby making the initial decohesion stages more energy-consuming. Given that both methods yield comparable fracture energies at this strain rate (Figure~\ref{fig:expanding_ring_results}b), the results confirm that the modified \acrshort{TSL} initially requires more energy to damage cohesive elements, ultimately slowing fragment formation. It underscores how the choice of traction-separation law influences fragmentation outcomes. As the strain rate increases, the damage levels in the cohesive elements rise, and the behavior converges to that of the standard linear \acrshort{TSL}, resulting in consistent fragment sizes between the two methods. We observe in Figure~\ref{fig:expanding_ring_results}b that both contact methods yield fracture energies that align in trend but exceed the ZMR lower bound, indicating that some energy is dissipated as diffuse damage ($d<1$) that does not necessarily lead to fragment formation. 

\begin{figure}[h!]
    \centering
    \includegraphics[scale=1]{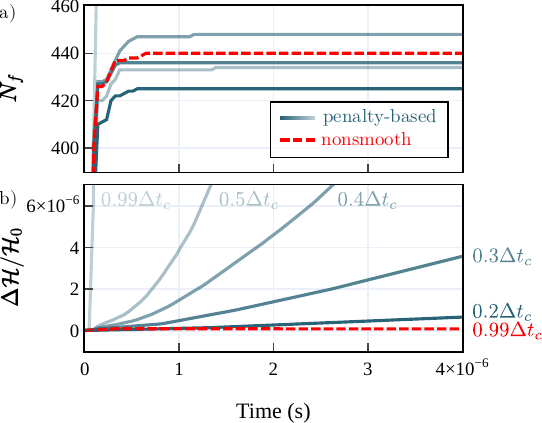}
    \caption{\textbf{a)} Sensitivity of the final fragment count to the time step size $\Delta t$ for the penalty-based method ($0.2 \Delta t_c \le \Delta t \le 0.99 \Delta t_c$) compared to the nonsmooth method ($\Delta t = 0.99 \Delta t_c$). \textbf{b)} Evolution of the total algorithmic energy variation over time for both methods and the varying time step sizes. The strain rate is set to $\hat{\dot{\varepsilon}}=1$ for both methods, and in Figure~\ref{fig:expanding_ring_results}, time steps were set at $0.2\Delta t_c$ and $0.99\Delta t_c$ respectively for penalty-based and nonsmooth approaches.}
    \label{fig:time_study_no_box}
\end{figure}

\medskip\noindent
As stated in Section~\ref{ss:impacting_bar_with_internal_damage}, the \acrshort{NSN} approach improves the overall stability of the simulation and allows the use of larger time steps, which offsets its computational overhead. In Figure~\ref{fig:time_study_no_box}, we present the final number of fragments and the evolution of the algorithmic energy \eqref{eq:algorithmic_energy} obtained for both contact methods as a function of the time step, for a strain rate of $\hat{\dot{\varepsilon}}=1$. While the nonsmooth method remains theoretically stable up to $\Delta t_c$, the penalty-based approach requires additional safety factors. We observe for the penalty-based approach that although the number of fragments is stable for time steps below $0.5 \Delta t_c$, maintaining energy conservation for long simulations would require time steps at least below $0.2 \Delta t_c$ for penalty-based methods. In that case, the effective overhead of the nonsmooth method would drop to a factor of $1.54$. This threshold of $\Delta t<0.2\Delta t_c$ is consistent with the results of \citep{ghesquiere2025stability}, although we see it does not guarantee long-term stability.

\medskip\noindent
In the next subsection, we explore a modified version of this benchmark in which the bar is confined to a box, resulting in more frequent and more intense contact interactions.

\subsection{Constrained expanding ring fragmentation}
\label{ss:constrained_expanding_ring_fragmentation}

This benchmark extends the previous one by introducing rigid walls that confine the bar's expansion. This modification significantly alters the fragmentation process by increasing the frequency and intensity of contact interactions. Fundamentally, the energy available for fracture switches from a local (fragment-scale) to a global (full bar-scale) energy budget.

\subsubsection{Energy budget}
We showed in Section~\ref{sssec:energy_budget_and_fragmentation_physics} that in free expansion, according to Grady, each fragment pays for its own creation using only its own local kinetic energy with a fragment size proportional to the strain rate as $\bar{s}_\text{free}\propto\dot{\varepsilon}^{-2/3}$~\eqref{eq:fragment_size_free}. The total dissipation $\mathcal{G}_\text{free}=(L/\bar{s}_\text{free})G_c$ results in an energy density ($\mathcal{G}_\text{free}/L$) that is entirely independent of the total bar length.

\medskip\noindent
When the expansion is confined within boundaries, such as a rigid box, the fragments cannot simply fly away; instead, they collide with the walls and one another. These interactions effectively make the global kinetic energy, which would otherwise be lost to rigid-body motion, available to drive additional cracking. In this confined state, the energy source is no longer restricted to the local domain of a single fragment, but is instead the integral of kinetic energy across the full length $L$ of the bar:
\begin{equation}
    \mathcal{K}_\text{global} = \int_{-L/2}^{L/2} \frac{1}{2} \rho (\dot{\varepsilon} x)^2 dx = \frac{\rho L^3 \dot{\varepsilon}^2}{24}.
\end{equation}
Assuming an idealized limit where all this available global energy is dissipated through the creation of $N$ fragments, the total fracture energy $\mathcal{G}_\text{confined}$ must equal $\mathcal{K}_\text{global}$. The resulting balance, $\mathcal{K}_\text{global}=(L/\bar{s}_\text{confined})G_c$, leads to a new characteristic fragment size:
\begin{equation}
    \bar{s}_\text{confined} = \frac{24G_c}{\rho L^2 \dot{\varepsilon}^2}.
\end{equation}
While the energy dissipated per unit length in free expansion is constant, the dissipated energy density in the confined case ($\mathcal{G}_\text{confined}/L$) scales quadratically with the bar length $L^2$. The magnitude of this effect is quantified by the ratio:
\begin{equation}
    r=\frac{\mathcal{K}_\text{global}}{\mathcal{K}_\text{local}}=\left(\frac{L}{\bar{s}_\text{free}}\right)^2
\end{equation}
This confinement forces the system to produce much smaller fragment sizes and dissipate more energy than the free-expansion baseline.

\subsubsection{Numerical implementation and results}
To capture this confinement behavior and evaluate these analytical limits, we employ the proposed semi-explicit \acrshort{NSN} scheme, in which contact is treated as fully elastic, i.e., with a coefficient of restitution $e=1$. We focus on a unit normalized strain rate ($\hat{\dot{\varepsilon}}=1$) that we choose at the limit of the dynamic regime for computational reasons: with higher strain rates, the available kinetic energy for fracture increases quadratically, requiring much finer meshes and millions of time steps to achieve convergence. Ideally, this strain rate should be higher. This choice defines a fixed base fragment size $\bar{s}_\text{free}$, and we study bars of varying lengths $L$ to explore different global-to-local energy ratios $r$. The box size $L_\text{box}$ is calibrated to ensure that wall collisions occur after the initial free-expansion fragmentation has concluded:
\begin{equation}
    L_\text{box} = L \left[1+\alpha_\text{box}\left(\frac{\sigma_c}{E} + \frac{\bar{s}_\text{free}\dot\varepsilon}{c}\right)\right].
\end{equation}
The term $\sigma_c/E$ in the $\alpha_\text{box}$ product accounts for the static elastic strain required to reach the cohesive strength $\sigma_c$. The second term represents the distance the bar's edges travel during the time required for an elastic wave to propagate through a fragment of size $\bar{s}_\text{free}$. The parameter $\alpha_\text{box}$ scales that spatial buffer. Because secondary fragmentation in this 1D benchmark arises from a sequence of spalling events, high confinement (small $\alpha_\text{box}$) prevents it; frequent impacts keep the fragments in compression, preventing the buildup of tensile stress required for spalling. Increasing the box size provides each fragment sufficient time and space for tensile waves to develop. To ensure a sufficiently developed spalling cascade, we set $\alpha_\text{box}=10^2$ for the rest of the analysis.

\begin{figure}[h!]
    \centering
    \includegraphics[scale=1]{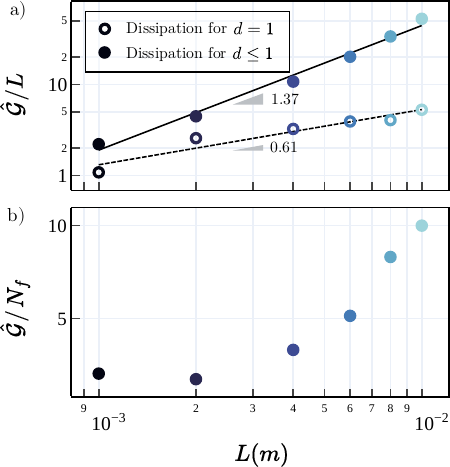}
    \caption{Effect of the domain length $L$ on the fracture energy. \textbf{a)} Scaling of the total normalized dissipated energy ($\hat{\mathcal{G}}/L\propto L^{1.37}$) compared to the energy associated with fully formed fracture surfaces ($\propto 1/\hat{s} \propto L^{0.61}$). While the latter accounts only for fully formed cracks ($d=1$), the total dissipation includes energy dissipated in partially damaged cohesive zones ($d<1$). \textbf{b)} Total fracture energy $\mathcal{G}$ normalized by the theoretical fracture energy required to create $N_f$ fragments, $N_f\times G_c$. From both plots, we see that the increase in the bar length $L$ leads to a significant increase in the total dissipated energy, which grows faster than the theoretical fracture energy associated with fully formed cracks. This discrepancy highlights the growing contribution of internal damage (partially damaged cohesive zones) to the overall energy dissipation as the bar length increases.}
    \label{fig:length_study_box}
\end{figure}

\medskip\noindent
The physical limitations of idealized confined fragmentation are synthesized in Figure~\ref{fig:length_study_box}. By analyzing the scaling exponents of the different energy sinks, we can partition them and identify why the system departs from the analytical $L^2$ limit of energy dissipation. We distinguish three distinct types of energy allocation based on the results in Figure~\ref{fig:length_study_box}a:
\begin{itemize}
    \item \textbf{Surface creation} ($\sim L^{0.61}$): This represents the energy strictly consumed by the creation of fully formed cracks ($d=1$), which scales with the inverse of the mean fragment size ($\propto 1/\hat{s}$). The low exponent indicates that, as the bar length increases, the creation of new surfaces becomes a smaller fraction of the total energy dissipation.
    \item \textbf{Internal damage} ($\sim L^{0.61}$ to $\sim L^{1.37}$): The gap between the surface creation energy ($0.61$) and the total dissipated energy scaling ($1.37$) represents energy consumed by ``internal damage''--- cohesive zones that have been activated ($d>0$) but fail to reach full separation ($d<1$). This phenomenon is explicitly captured in Figure~\ref{fig:length_study_box}b, which plots the ratio $\hat{\mathcal{G}}/N_f$ (total energy dissipated per fully formed fragment). The fact that this ratio climbs from approximately $2$ at small scales to $10$ at larger scales confirms that as the bar length increases, the fragment creation process becomes less efficient.
    \item \textbf{Residual trapped energy} ($\sim L^{1.37}$ to $\sim L^{2}$): The final gap between the total dissipated energy and the theoretical limit represents the energy that the system is unable to dissipate through any fracture mechanism. This energy is retained within the fragments as rigid-body motion (kinetic energy) and internal vibrations (kinetic and strain energy). Because a stress threshold governs fragmentation, this energy remains trapped when stresses generated from impacts or internal wave reflections fail to reach the cohesive strength $\sigma_c$ required to initiate or advance further fracture.
\end{itemize}

\medskip\noindent
The $L^{1.37}$ scaling of energy dissipation is reasonably close compared with the expected $L^2$ scaling. In addition to the energy being trapped in the system rather than dissipated, the strain rate we chose at the limit of the dynamic regime, $\hat{\dot{\varepsilon}}=1$, can explain the observed gap.

\medskip\noindent
The results discussed thus far were obtained under the assumption of perfectly elastic contact ($e=1$). However, the introduction of contact dissipation ($e<1$) allows the model to account for energy lost during collisions. Leveraging the proposed \acrshort{NSN} scheme, we explore these effects in the following section.

\subsection{Effect of contact dissipation on secondary fragmentation}

\begin{figure*}[h!]
    \centering
    \includegraphics[scale=1]{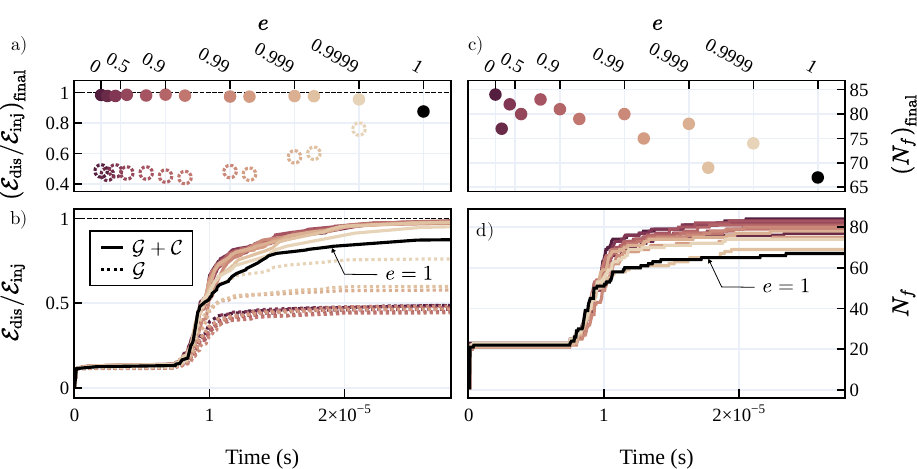}
    \caption{On all subfigures, the color scale shows different values of $e\in[0, 0.9999]$ while black shows the case with $e=1$. \textbf{a)} Final values of the normalized fracture energy $\mathcal{G}/\mathcal{E}_\text{inj}$ (dashed circles) and total dissipated energy summing fracture and contact, $(\mathcal{G}+\mathcal{C})/\mathcal{E}_\text{inj}$ (solid circles) as a function of $e$. \textbf{b)} Evolution in time of these energies for different restitution coefficients $e$. Dashed and solid lines follow the same logic as \textbf{a)}. \textbf{c)} Final number of fragments $N_f$ as a function of the restitution coefficient $e$. \textbf{d)} Evolution in time of the number of fragments for different restitution coefficients $e$. The results show that while the total dissipated energy $\mathcal{G}+\mathcal{C}$ is essentially independent of $e$ for any inelastic contact ($e<1$), the number of fragments increases as $e$ decreases, suggesting that contact dissipation promotes damage localization and the formation of fully separated fragments.}
    \label{fig:restitution_study}
\end{figure*}

\medskip\noindent
When using perfectly elastic contact in confined environments, once fracture saturates (i.e., no new cohesive surfaces can form), the remaining kinetic energy is conserved indefinitely. Introducing dissipative contact ($e<1$) allows the model to dissipate energy during collisions, prompting the question of how contact dissipation affects the partitioning of energy dissipation and, consequently, secondary fragmentation. We study how the number of fragments and the energies dissipated by fracture and contact vary with the restitution coefficient $e\in[0,1]$. The studied bar has a length $L=5\cdot10^{-3}$ m, its free-expansion strain rate is set to $\hat{\dot{\varepsilon}}=1$, and it is confined within a box whose size is scaled by $\alpha_\text{box}=100$. Figure~\ref{fig:restitution_study}a reports the dissipated energy partition evolution, now split between fracture dissipation $\mathcal{G}$ (dashed lines) and contact dissipation $\mathcal{C}$, summed with $\mathcal{G}$ (solid lines). Both are normalized by the total injected energy, 
\[ 
    \mathcal{E}_\text{inj}=\mathcal{E}_0+\mathcal{W}_\text{ext}(t_\text{final}), 
\] 
i.e., the sum of the initial energy (kinetic and strain energy) and the external work. The bottom plot shows their respective time histories, while the top one shows the final values. With contact dissipation ($e<1$), the fracture energy $\mathcal{G}$ drops from about $88\%$ of $\mathcal{E}_\text{inj}$ in the purely elastic case ($e=1$) to roughly $45\%$ for $e<0.999$. Between $e=0.9999$ and $e=1$, the fracture energy increases gradually to the value of the purely elastic case. However, the total dissipated energy $\mathcal{G}+\mathcal{C}$ grows faster than in the perfectly elastic case and reaches about $98\%$ of $\mathcal{E}_\text{inj}$ over the simulated time. Remarkably, for any $e<1$, the curves for $\mathcal{G}+\mathcal{C}$ collapse approximately onto a single evolution, suggesting that the total dissipated energy is only slightly dependent on $e$ as soon as contact is inelastic and that contact zones are numerous.

\begin{figure}[h!]
    \centering
    \includegraphics[scale=1]{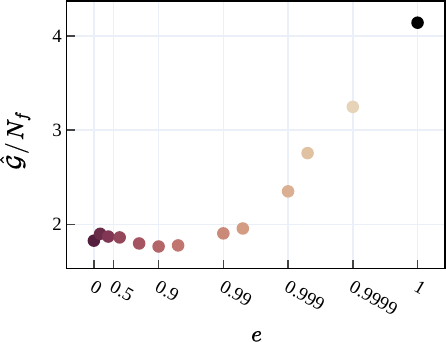}
    \caption{Actual total fracture energy $\mathcal{G}$ normalized by the theoretical fracture energy required to create $N_f$ fragments, $N_f\times G_c$, as a function of the restitution coefficient $e$. The efficiency of the fragmentation process increases as $e$ decreases, as less energy is required to create more fragments.}
    \label{fig:restitution_efficiency}
\end{figure}

\medskip\noindent 
Counterintuitively, a strong reduction in fracture dissipation does not lead to fewer fragments. As shown in Figure~\ref{fig:restitution_study}b, systems incorporating contact dissipation actually produce more fragments than the perfectly elastic baseline, with the total fragment count steadily increasing as the restitution coefficient decreases. This indicates that contact dissipation makes the fragmentation process significantly more efficient at localizing damage and driving cracks to full separation. Since the theoretical energy required to form a single new fragment is exactly the fracture energy $G_c$, we can quantify the overall efficiency of the fragmentation using the ratio $\hat{\mathcal{G}}/N_f$. It normalizes the actual total fracture energy by the theoretical fracture energy required to produce $N_f$ fragments. As illustrated in Figure~\ref{fig:restitution_efficiency}, this ratio decreases drastically---from $\sim4$ in the perfectly elastic case down to $\sim1.8$ for $e<0.999$. Contact dissipation therefore promotes a fragmentation regime that is more than twice as efficient by suppressing energy lost to partially damaged cohesive zones ($d<1$).

\medskip\noindent
To rationalize this contradiction, we hypothesize that contact dissipation modifies damage localization by filtering wave propagation. In the perfectly elastic case ($e=1$), stress waves reflect off boundaries and crack faces without attenuation, maintaining a highly vibratory stress field along the bar. This environment promotes diffuse damage growth in many cohesive elements and favors the formation of extended damaged zones rather than clean separations. For $e<1$, contact dissipation effectively acts as a low-pass filter on the stress field, systematically suppressing high-frequency oscillations and keeping cohesive gaps small. Crucially, this damping mechanism is driven by contact; therefore, its effect scales directly with the number of contact zones. As the material accumulates damage and generates more cohesive interfaces, the increased frequency of contact events accelerates wave attenuation. By filtering out such wave interference, this process allows tensile stresses to localize more stably, promoting clean, localized fractures (see supplemental movies).

\medskip\noindent
However, this phenomenon should not motivate a choice for $e$ without a more detailed understanding of the underlying physics of such a coefficient. In real materials, the restitution coefficient is not constant but a complex function of local dissipation, dependent on the relative contact velocity, surface roughness, and material properties. Future work should aim to characterize this relationship to inform more physically accurate choices for $e$ in fragmentation simulations.

%% file: 05_conclusion.tex
\section{Conclusion}
\label{sec:conclusion}

In this work, we adapted and validated a novel nonsmooth explicit-implicit time-integration scheme, the \acrfull{NSN} method ($\beta=0, \gamma=\frac{1}{2}$), designed specifically for simulating dynamic fragmentation with nonsmooth unilateral contact. Following \cite{chen2013nonsmooth}, the proposed scheme combines the computational efficiency of explicit time integration for bulk dynamics and cohesive fracture with the robustness of the \acrshort{NSCD} method for resolving complex contact interactions, while maintaining second-order accuracy during free-flight phases. Through a series of benchmark problems, the \acrshort{NSN} scheme demonstrated excellent performance compared to reference solutions and existing nonsmooth methods. For constant acceleration scenarios, such as the bouncing ball test, the method exhibited accuracy comparable to both the implicit \acrfull{MJ} scheme ($\theta=\frac{1}{2}$) and the semi-explicit \acrfull{CDL} scheme. For more complex impact dynamics, such as a bar striking a rigid wall, the \acrshort{NSN} scheme maintained its accuracy (second-order on smooth terms, first order on nonsmooth terms)---matching the \acrshort{CDL} scheme and outperforming the degrading \acrshort{MJ} scheme---while demonstrating superior robustness to variations in the coefficient of restitution.

\medskip\noindent
To bridge our numerical framework with previous studies detailing the instabilities of penalty-based contact formulations with an explicit approach~\cite{ghesquiere2025stability}, we analyzed a variant of the impacting bar featuring internal cohesive damage ($d<1$). Our results confirm that the \acrshort{NSN} approach effectively represents the converged solution of the penalty method. Compared with the highest penalty stiffnesses tested, the \acrshort{NSN} scheme achieved similar displacement accuracy, improved velocity accuracy, and machine-precision energy conservation. Crucially, the nonsmooth approach eliminated the numerical instabilities inherent to penalty-based methods and exhibited exceptional robustness to time-step variations. While the \acrshort{NSN} scheme incurs a per-step computational cost roughly five times that of a purely explicit scheme with penalty-based contact, its improved stability permits the use of significantly larger time steps, ultimately resulting in a substantial reduction in the overall computational cost for a given accuracy threshold.

\medskip\noindent
Finally, we applied the proposed method to a highly dynamic benchmark: the one-dimensional fragmentation of a bar under a constant strain rate. For free expansion, the method accurately reproduced the expected ZMR predictions for fragment size \cite{zhou2005cohesive}, confirming that the new contact treatment does not artificially influence the results. When the bar's expansion is confined to rigid boundaries, the numerical model successfully captures a fundamental shift in fragmentation physics. Confinement shifts the energy budget available for fracture from the local kinetic energy of individual fragments to the global kinetic energy of the entire system, leading to a drastic increase in fracture energy dissipation and a corresponding reduction in the mean fragment size. Furthermore, the robust enforcement of nonsmooth contact allowed us to uncover counterintuitive dynamics regarding the role of contact dissipation during this confined cascade of spalling events. We demonstrated that decreasing the restitution coefficient $e$ reduces the energy dissipated for fracture but increases the final number of fragments. Introducing contact dissipation helps localize damage and drives cracks to full separation, thereby preventing the system from dissipating energy in distributed, partial damage zones ($d<1$).

\medskip\noindent
The \acrshort{NSN} scheme is also valid for two- and three-dimensional problems. Future work will implement this method in the fully parallelized, high-performance environment of the open-source code \texttt{Akantu} \citep{richart2024akantu}.

%% file: 06_appendix.tex
\section{Symmetry and positive semidefiniteness of the modified Delassus operator.}
\label{appendix:spd}

To solve the \acrshort{LCP}~\eqref{eq:lcp} as a constrained quadratic program~\eqref{eq:qp}, the modified Delassus operator must be \acrfull{SPSD}. We show here that this is indeed the case under the usual explicit dynamics time-step condition.
Recall that
\[
\mathbf{W}' = \mathbf{H}^{\mathcal{A}}\mathbf{M}^{-1}\left(\mathbf{H}^{\mathcal{A}}\right)^{\!\top} - \tfrac{\Delta t^{2}}{4}\,\mathbf{H}^{\mathcal{A}}\mathbf{M}^{-1}\mathbf{K}\mathbf{M}^{-1}\left(\mathbf{H}^{\mathcal{A}}\right)^{\!\top}.
\]
We assume throughout that the mass matrix $\mathbf{M}$ is \acrfull{SPD} and the stiffness matrix $\mathbf{K}$ is \acrshort{SPSD}.

\paragraph{Symmetry.} Since $\mathbf{M}^{-1}$ is symmetric and $\mathbf{K}$ is symmetric,
\[
\left(\mathbf{H}^{\mathcal{A}}\mathbf{M}^{-1}\left(\mathbf{H}^{\mathcal{A}}\right)^{\!\top}\right)^{\!\top}
= \mathbf{H}^{\mathcal{A}}\mathbf{M}^{-1}\left(\mathbf{H}^{\mathcal{A}}\right)^{\!\top},
\]
and, by the same argument,
\[
\left(\mathbf{H}^{\mathcal{A}}\mathbf{M}^{-1}\mathbf{K}\mathbf{M}^{-1}\left(\mathbf{H}^{\mathcal{A}}\right)^{\!\top}\right)^{\!\top}
= \mathbf{H}^{\mathcal{A}}\mathbf{M}^{-1}\mathbf{K}\mathbf{M}^{-1}\left(\mathbf{H}^{\mathcal{A}}\right)^{\!\top}.
\]
Hence $\mathbf{W}'$ is symmetric.

\paragraph{Positive semidefiniteness under the standard time-step condition.}
For any $\mathbf{x}$, set $\mathbf{y} := \mathbf{M}^{-1/2}\left(\mathbf{H}^{\mathcal{A}}\right)^{\!\top}\! \mathbf{x}$. Then
\[
\mathbf{x}^\top \mathbf{W}' \mathbf{x}
= \|\mathbf{y}\|^{2}
\;-\; \tfrac{\Delta t^{2}}{4}\,
\mathbf{y}^\top\bigl(\mathbf{M}^{-1/2}\mathbf{K}\mathbf{M}^{-1/2}\bigr)\mathbf{y}.
\]
Let $\lambda_{\max}$ be the largest eigenvalue of the SPSD matrix
$\mathbf{M}^{-1/2}\mathbf{K}\mathbf{M}^{-1/2}$ and define $\omega_{\max}:=\sqrt{\lambda_{\max}}$. Then
\[
\mathbf{x}^\top \mathbf{W}' \mathbf{x}
\;\ge\; \Bigl(1 - \tfrac{\Delta t^{2}}{4}\lambda_{\max}\Bigr)\,\|\mathbf{y}\|^{2}
= \Bigl(1 - \tfrac{\Delta t^{2}}{4}\omega_{\max}^{2}\Bigr)\,\|\mathbf{y}\|^{2}.
\]
Therefore,
\[
\mathbf{W}'\succeq 0 \quad \text{whenever} \quad \Delta t \le \frac{2}{\omega_{\max}}.
\]
This is a \emph{sufficient (not necessary)} time-step condition that ensures the convexity of the \acrshort{QP}; it coincides with the standard central-difference bound. In particular, $\mathbf{W}'$ may remain SPSD for larger $\Delta t$ if the contact subspace does not excite the stiffest mode (the estimate uses the global spectral radius).

\section{Equivalence of implicit Moreau-Jean and semi-explicit nonsmooth Newmark under constant acceleration}
\label{appendix:bouncing_ball}

In this section, we demonstrate how the implicit \acrfull{MJ} and the explicit \acrfull{NSN} ($\beta=0$, $\gamma=\tfrac12$) schemes produce identical trajectories for the bouncing ball problem when the gravitational acceleration $g$ is constant and the time step $\Delta t$ is chosen such that impact detection is synchronized. We assume a constant acceleration $a(t) = -g$, a time step $\Delta t$, and a restitution coefficient $e$. The state at time step $n$ is given by position $u_n$ and velocity $v_n$.

\subsection{Smooth phase (free flight)}
Between impacts, the motion is governed purely by gravity.

\paragraph{Moreau-Jean Scheme:}
The velocity update is explicit Euler:
\begin{equation}
    v_{n+1} = v_n - g \Delta t
    \label{eq:mj_vel_ff}
\end{equation}
The position update uses the trapezoidal rule:
\begin{equation}
    u_{n+1} = u_n + \frac{\Delta t}{2} (v_n + v_{n+1})
    \label{eq:mj_pos_ff}
\end{equation}
Substituting \eqref{eq:mj_vel_ff} into \eqref{eq:mj_pos_ff} gives:
\begin{equation}
    u_{n+1}^\text{MJ} = u_n + \frac{\Delta t}{2} (v_n + v_n - g \Delta t) = u_n + v_n \Delta t - \frac{1}{2} g \Delta t^2
    \label{eq:mj_free_flight}
\end{equation}

\paragraph{Nonsmooth Newmark scheme:}
The standard explicit Newmark predictor for displacement with $\beta=0$ and $\gamma=\tfrac12$ (constant acceleration) is:
\begin{equation}
    u_{n+1}^\text{NSN} = u_n + v_n \Delta t + \frac{1}{2} a_n \Delta t^2 = u_n + v_n \Delta t - \frac{1}{2} g \Delta t^2
    \label{eq:nsn_pos_ff}
\end{equation}
Comparing \eqref{eq:mj_free_flight} and \eqref{eq:nsn_pos_ff}, the kinematic updates for the smooth phase are identical.

\subsection{Impact Phase}
By using the respective gap prediction of \acrshort{MJ} and \acrfull{NSN} schemes, we assume both predictions detect a contact at step $n+1$. The post-impact velocity is defined by the restitution law $v_{n+1} = -e v_n$.

\paragraph{Moreau-Jean scheme:}
The position update remains a trapezoidal average of velocities:
\begin{equation}
    u_{n+1}^\text{MJ} = u_n + \frac{\Delta t}{2} (v_n + v_{n+1}) = u_n + \frac{\Delta t}{2} v_n (1 - e)
    \label{eq:mj_pos_impact}
\end{equation}

\paragraph{Nonsmooth Newmark scheme:}
The NSN scheme first computes the ``smooth'' predictor values, denoted by $(\;\widetilde\cdot\;)$:
\begin{align}
    \widetilde{u}_{n+1} &= u_n + v_n \Delta t - \frac{1}{2} g \Delta t^2 \\
    \widetilde{v}_{n+1} &= v_n - g \Delta t
\end{align}
The velocity jump $\widehat{v}_{n+1}$ required to satisfy the restitution law is the difference between the target velocity and the predictor velocity:
\begin{equation}
    \widehat{v}_{n+1} = v_{n+1} - \widetilde{v}_{n+1} = (-e v_n) - (v_n - g \Delta t)
\end{equation}
The final position is updated using the Newmark correction term $\frac{\Delta t}{2} \widehat{v}_{n+1}$:
\begin{equation}
    u_{n+1}^\text{NSN} = \widetilde{u}_{n+1} + \frac{\Delta t}{2} \widehat{v}_{n+1}
\end{equation}
Substituting the expressions for $\widetilde{u}_{n+1}$ and $\widehat{v}_{n+1}$:
\begin{equation}
    u_{n+1}^\text{NSN} = \left( u_n + v_n \Delta t - \frac{1}{2} g \Delta t^2 \right) + \frac{\Delta t}{2} \left( -e v_n - v_n + g \Delta t \right)
\end{equation}
With the acceleration terms $\frac{1}{2} g \Delta t^2$ canceling out exactly, expanding the correction term simplifies to:
\begin{equation}
    u_{n+1}^\text{NSN} = u_n + \frac{\Delta t}{2} v_n (1 - e)
\end{equation}
Comparing this result with Eq.~\eqref{eq:mj_pos_impact}, we see that $u_{n+1}^\text{MJ} = u_{n+1}^\text{NSN}$. Thus, provided that the contact-detection logic is triggered at the same step, both schemes yield algebraically identical updates.

\section{Expanding ring simulation parameters}
\label{appendix:expanding_ring_simulation_parameters}

The bar is discretized into $N_e$ linear (P1) bar elements. To improve numerical convergence, the length of each element $h_e = L/N_e$ is randomized uniformly between $\pm 0.4\times h_e$ as recommended by\;\cite{molinari2007cohesive}. To introduce strength variation in the bar, we randomly assign $N_{\text{def}}$ defect locations along the bar. At these sites, the cohesive strength is drawn from a uniform distribution $\sigma_{c,\text{def}}\in[0.98, 1]\sigma_c$. The choice of $N_{\text{def}}$ is guided by two main criteria:
\begin{itemize}
  \item \textbf{Mesh independence:} By using a fixed number of defect sites rather than varying the strength of every potential interface along the mesh, we prevent the fragment size distribution from becoming mesh-dependent. Without this constraint, refining the mesh would artificially increase the probability of encountering a ``weak link'' per unit length, potentially shifting the failure mode from localized cracking to non-physical distributed damage~\citep{zhou2005cohesive,levy2010dynamic}.
  \item \textbf{Defect spacing:} The number of defects, $N_{\text{def}}$, is selected so that the average spacing between defects is significantly smaller than the expected fragment size at a given strain rate. It ensures that the available defect sites are sufficient to capture the fragmentation process without limiting the number of fragments formed. Consequently, the choice of $N_{\text{def}}$ is also dependent on the imposed strain rate.
\end{itemize}

\medskip\noindent
The domain length $L$, the mesh density $\rho_e=N_e/L$ and defect density $\rho_{\text{def}}=N_{\text{def}}/L$ are termed converged at a given ($\hat{\dot{\varepsilon}}$) when further refinement of $\rho_e$ (and, respectively, of $\rho_{\text{def}}$ at fixed $\rho_e/\rho_{\text{def}}$) produces negligible changes in the number of fragments and in the energy dissipation. To maintain a statistical significance of the results and numerical convergence across the range of strain rates, they are adjusted as follows:
\begin{itemize}
  \item \textbf{Quasi-static regime ($\hat{\dot{\varepsilon}} \leq 1$):} The domain length is set to $L=10^{-1} \text{m}$, which serves as a \acrfull{RVE} for the lowest strain rate ($\hat{\dot{\varepsilon}}=10^{-3}$) where fragments are the largest. Specifically, it allows us to capture $\sim100$ fragments at this strain rate. The mesh density is converged at $\hat{\dot{\varepsilon}}=1$, where fragment sizes are the smallest in that regime, resulting in $\rho_e=5\times10^5 \text{m}^{-1}$. Similarly, the defect density is converged at that same strain rate to a value of $\rho_\text{def}=10^5~\text{m}^{-1}$. These values ($L$, $\rho_e$, $\rho_\text{def}$) are held constant for all $\hat{\dot{\varepsilon}} \leq 1$.
  \item \textbf{Dynamic regime ($\hat{\dot{\varepsilon}} > 1$):} In the dynamic regime, Grady‑type scaling and numerical studies indicate that the characteristic fragment size decreases as $\hat{\dot{\varepsilon}}^{-2/3}$. To retain a comparable number of fragments for each strain rate while controlling computational cost, we scale the domain length as $L\propto \hat{\dot{\varepsilon}}^{-2/3}$. At the same time, we increase the mesh density as $\rho_e\propto \hat{\dot{\varepsilon}}^{2/3}$ so that the number of elements per fragment remains approximately constant. Finally, we keep the element‑to‑defect ratio constant at $\rho_e/\rho_{\text{def}}=5$. This strategy preserves accuracy and statistical representativity across strain rates while limiting the total number of degrees of freedom.
\end{itemize}

\medskip\noindent
These simulation parameters are summarized in Table~\ref{tbl:simulations}. 

\begin{table*}[h!]
  \centering
  \begin{tabular}{l c c c c c c c}
    \toprule
    & \multicolumn{7}{c}{\textbf{Imposed strain rate} $\hat{\dot{\varepsilon}}$} \\
    \cmidrule(lr){2-8}
                                                  & $10^{-3}$           & $10^{-2}$           & $10^{-1}$           & $1$                 & $10$                  & $10^{2}$              & $10^{3}$            \\
    \midrule
    Bar length, \(L\) [m]                         & $10^{-1}$           & $10^{-1}$           & $10^{-1}$           & $10^{-1}$           & $2.15\times10^{-2}$   & $4.64\times10^{-3}$   & $10^{-3}$           \\
    Mesh density, \(\rho_e\) [m\(^{-1}\)]         & $5.0\times10^{5}$   & $5.0\times10^{5}$   & $5.0\times10^{5}$   & $5.0\times10^{5}$   & $2.32\times10^{6}$    & $1.08\times10^{7}$    & $5.0\times10^{7}$   \\
    Defect density, \(\rho_{\text{def}}\) [m\(^{-1}\)]   & $1.0\times10^{5}$   & $1.0\times10^{5}$   & $1.0\times10^{5}$   & $1.0\times10^{5}$   & $4.64\times10^{5}$    & $2.16\times10^{6}$    & $1.0\times10^{7}$   \\
    \bottomrule
  \end{tabular}
  \caption{Simulation parameters for each imposed strain rate.}
  \label{tbl:simulations}
\end{table*}